\documentclass[useAMS,usenatbib,usegraphicx]{mn2e}
\usepackage{amsmath}
\usepackage{amssymb}
\usepackage{booktabs}
\usepackage{times}
\usepackage{aas_macros} 
\usepackage[usenames]{color}

\title[A rare triple-mode RRd star]{EPIC\,201585823, a rare triple-mode RR\,Lyrae star discovered in K2 mission data}

\author[Kurtz et al.]
{Donald W. Kurtz$^1$, Dominic M. Bowman$^1$,  Simon J. Ebo$^1$, Pawe{\l} Moskalik$^2$,   \newauthor{Rasmus Handberg$^3$ and Mikkel N. Lund$^3$ }\\
$^1$Jeremiah Horrocks Institute, University of Central Lancashire, Preston PR1 2HE, UK\\
$^2$Copernicus Astronomical Center, ul. Bartycka 18, PL-00-716, Warsaw, Poland\\
$^3$Stellar Astrophysics Centre (SAC), Department of Physics and Astronomy, Aarhus University,
Ny Munkegade 120, DK-8000 Aarhus C, Denmark\\
}

\begin{document}

\maketitle

\begin{abstract}
We have discovered a new, rare triple-mode RR\,Lyr star, EPIC\,201585823, in the {\it Kepler} K2 mission Campaign 1 data. This star pulsates primarily in the fundamental and first-overtone radial modes, and, in addition, a third nonradial mode. The ratio of the period of the nonradial mode to that of the first-overtone radial mode, 0.616285, is remarkably similar to that seen in 11 other triple-mode RR\,Lyr stars, and in 260 RRc stars observed in the Galactic Bulge. This systematic character promises new constraints on RR\,Lyr star models. We detected subharmonics of the nonradial mode frequency, which are a signature of period doubling of this oscillation; we note that this phenomenon is ubiquitous in RRc and RRd stars observed from space, and from ground with sufficient precision. The nonradial mode and subharmonic frequencies are not constant in frequency or in amplitude. The amplitude spectrum of EPIC\,201585823 is dominated by many combination frequencies among the three interacting pulsation mode frequencies. Inspection of the phase relationships of the combination frequencies in a phasor plot explains the `upward' shape of the light curve. We also found that raw data with custom masks encompassing all pixels with significant signal for the star, but without correction for pointing changes, is best for frequency analysis of this star, and, by implication, other RR\,Lyr stars observed by the K2 mission.  We compare several pipeline reductions of the K2 mission data for this star.
\end{abstract}


\section{Introduction}
\label{sec:intro}

RR\,Lyrae stars were first noted in the 1890s by Solon Bailey, Wilhelmina Fleming and Edward Pickering. At first, they were known as cluster variables because so many were found in globular clusters. Those variables, along with globular cluster Cepheids, allowed Harlow Shapley, in his `Studies of Magnitudes in Star Clusters' series of papers beginning in 1916,  to measure distances to the globular clusters and show that the Sun lies far from the centre of the Milky Way, using Henrietta Leavitt's now-famous 1912 period-luminosity relation \citep{leavitt1912}, along with Ejnar Hertzsprung's 1913 calibration using statistical parallaxes of 13 Cepheids \citep{hertzsprung1913}. In the early decades of the 20$^{\rm th}$ century the RR\,Lyr stars had made their mark as distance indicators, and it is that for which they are still best known. 

Solon Bailey first classified the light curves of RR\,Lyr stars phenomenologically into three sub-classes, `a', `b' and `c', in his 252-page magnum opus on over 500 variable stars in $\omega$\,Cen \citep{bailey1902}. Bailey's classes `a' and `b' stars are now consolidated into a class known as RRab stars that pulsate in the fundamental radial mode and have strongly non-sinusoidal light curves; his class `c' stars are now known as RRc stars, which are variables that pulsate in the first-overtone radial mode with more sinusoidal variations. 

Edward Pickering, the director of the Harvard College Observatory, noted in the preface to Bailey's work: `Nearly all of these stars appear to vary with perfect regularity so that the period can be determined, in some cases, within a fraction of a second'. Yet it was a mere 5 years later that \citet{blazkho1907} first noted modulation of the light curve of the RR\,Lyr star RW\,Dra. He reported that the period varied between 10$^{\rm h}$\,35$^{\rm m}$\,36$^{\rm s}$ and 10$^{\rm h}$\,40$^{\rm m}$\,02$^{\rm s}$ in the time span of 41.6\,d. So much for Pickering's `perfect' regularity! Modulation in RR\,Lyr light curves is common and now known as the Blazhko effect, a term first used by \citet{tsessivich1953}. (`Blazhko' is now the spelling universally used, rather than the original spelling of `Bla\v zko'.) This effect is not yet fully understood.

\citet{jerz1977} discovered that AQ\,Leo is an RR\,Lyr star pulsating in two modes that are nonlinearly interacting, generating many combination frequencies. That was the first `double mode' RR\,Lyr star discovered, hence can be considered to be the prototype of the class. Since then, nearly 2000 more such stars have been found \citep{moskalik2014}, and they are now known as RRd stars. The `d' in RRd fits nicely in sequence with Bailey's  `a', `b' and `c', and fortuitously it also can be construed to mean `d' for double mode. Typically, the double mode RR\,Lyr stars pulsate in the fundamental and first-overtone radial modes with a period ratio close to 0.74 for most such stars and usually the first-overtone mode has the higher amplitude \citep{moskalik2013}. 

More recently, other RRc stars have been found pulsating in two modes, the first-overtone radial mode and a nonradial mode with frequency between the third and fourth overtone radial modes with a period ratio with respect to the dominant first-overtone radial mode near to 0.61 (the range is $0.595 - 0.634$). While these stars are relatively rare compared to all known RR\,Lyr stars, many are now known. The first RRc stars with the 0.61 period ratio were found by \citet{olech2009}, who found two unambiguous cases in $\omega$\,Cen, V19 and V105, which they announced to be the first members of a new class of double mode RR Lyr stars. A few years later \citet{moskalik2015} listed only 19 such stars (their Table 8); \citet{jurcsik2015} added 14 more (their Table 3) in a new study of RR Lyr stars in M3 (which is one of the first globular clusters studied for RR Lyr stars in the first decade of the 1900s by the Harvard group).

The number has now exploded dramatically with 145 such RRc stars discovered by \citet{netzel2015a} in the Galactic Bulge using OGLE-III data\footnote{OGLE is the Optical Gravitational Lensing Experiment studying the Galactic Bulge and the Magellanic Clouds. For general information on the OGLE-III and OGLE-IV surveys, see \citet{udalski2008}; \citet{udalski2015}; for specific studies of RR\,Lyr stars in the Galactic Bulge, see \citet{soszynski2011}; \citet{soszynski2014}.},
 although this represents only 3 per cent of the sample in that study. In another study of Galactic Bulge RR Lyr stars using OGLE-IV data$^1$, \citet{netzel2015b} analysed 485 RRc stars. They found nonradial modes with the period ratio near 0.61 in 131 RRc stars (115 of which are new discoveries). The occurrence rate is 27 per cent. The OGLE-IV data have a much higher data density than the OGLE-III data, which results in lower noise in the amplitude spectra; thus, as expected, there is a higher occurrence rate in the OGLE-IV data. Combining results of OGLE-IV and OGLE-III, there are now 260 RRc stars known in the Galactic Bulge with the nonradial mode with the 0.61 period ratio.

While RRc stars pulsating in the radial first-overtone mode and a nonradial mode with period ratio near to 0.61 are no longer rare, triple-mode RR\,Lyr stars are still rare. \citet{gruberbauer2007} observed the prototype RRd star, AQ\,Leo, with the MOST satellite for 34.4\,d and found, in addition to a large number of nonlinear combination frequencies, evidence for other, non-combination frequencies that they conjectured may be associated with additional modes. We can now see that they found a nonradial mode with the 0.61 period ratio with respect to the first-overtone radial mode. \citet{chadid2012} then found triple-mode pulsation in the RRd star CoRoT\,101368812, where the dominant mode frequencies have a ratio of 0.745, typical of fundamental and first-overtone radial modes in RRd stars, plus a third frequency belonging to a nonradial mode with a period ratio of 0.61 with respect to the first-overtone radial mode frequency. Two more such stars were found in K2 data by \citet{molnar2015}; three were found in the Galactic Bulge by \citet{netzel2015a} and \citet{smolecetal2015a}; and four were found in M3 by \citet{jurcsik2015}. Those 11, plus our discovery of triple-mode pulsation in EPIC\,201585823, brings to 12 the known triple-mode RR\,Lyr stars with the 0.61 period ratio. We list them with references in Section\,\ref{sec:additional}.

Yet another triple-mode RR\,Lyr variable was discovered by \citet{smolecetal2015b} in the OGLE-IV photometry of the Galactic Bulge. This star, OGLE-BLG-RRLYR-24137, pulsates in the radial fundamental mode, the radial first-overtone mode and a third mode with a period ratio of 0.686 to the radial first overtone.   This period ratio is very different from the 0.61 ratio observed in the RR\,Lyr stars discussed above. Thus, OGLE-BLG-RRLYR-24137 is a (currently) unique object, not similar to any other known RR\,Lyr pulsator. It will, therefore, not be discussed further in this paper.

\citet{moskalik1990} showed theoretically that any half-integer resonance (e.g., 3:2, 5:2, etc.) between two pulsation modes can cause `period doubling', a characteristic alternating of the amplitudes of the maxima or minima in the light curve. These alternations are represented in the Fourier spectrum by subharmonics, that is by peaks at $(n+\frac{1}{2})f$, where $n$ is integer and $f$ is the frequency of the period-doubled mode. Pulsations with such characteristics have been known for decades in the RV\,Tauri variables (for a review see, e.g., \citealt{wallerstein2002}). They were also predicted by \citet{buchler1992}, then discovered by \citet{smolec2012} in at least one BL\,Herculis star, BLG184.7\,133264 (BL\,Her stars are a short period subgroup of the Population\,II Cepheids). 

In the RR Lyrae stars, period doubling was first found (somewhat unexpectedly) by \citet{szabo2010}, who detected it in RR\,Lyr itself and in two other RRab variables observed by the {\it Kepler} space telescope. From hydrodynamic models, \citet{kollath2011}  then traced the period doubling in those stars to a 9:2 resonance between the ninth radial overtone and the fundamental radial mode. Moskalik et al. (2015) found period doubling of the nonradial 0.61 period ratio modes in 4 RRc stars observed by {\it Kepler}. They also noted from the literature a similar period doubling in 2 other RRc stars and 4 RRd stars observed from space (see their Table 8), showing that this phenomenon is ubiquitous. Although of very low amplitude, the period doubling subharmonics can also be detected in ground-based data of sufficiently good quality \citep{netzel2015b}. Not surprisingly then, we have found subharmonics also in EPIC\,201585823, which we discuss in Sections\,\ref{sec:additional} and \ref{sec:subharmonics}.

Multi-mode pulsation promises better asteroseismic information on the stellar structure of RR\,Lyr stars, hence deeper understanding both of stellar pulsation and of the interior properties of these important distance indicators. Asteroseismology depends on modelling identified mode frequencies \citep{aertsetal2010}, hence more frequencies give better constraints on models. In this paper we announce the discovery of a new triple-mode RR\,Lyr star, EPIC\,201585823, that is remarkably similar to 11 other triple-mode RR\,Lyr stars. We also compare and discuss several data reduction pipelines and other rectifications of {\it Kepler}  K2 mission data. 

\section{The triple mode RR\,Lyr star EPIC\,201585823}

\subsection{Data}
\label{sec:data}

The data for EPIC\,201585823 were obtained during Campaign 1 (C1) of the {\it Kepler} two-reaction-wheel extended mission, known as the K2 mission \citep{howell2014}. Several independent reductions and rectifications of the data are available publicly, and we compare these in Section\,\ref{sec:pipelines} below.  The K2 mission observes fields in the ecliptic plane where the two operational reaction wheels, balanced against solar radiation pressure, provide some pointing stability. Nevertheless, there is drift in the position of the satellite that must be corrected with thruster bursts on time scales of 5.9\,h (or multiples of that) and there are reaction wheel resaturations every 2\,d (\citealt{howell2014}; \citealt{vandenburg2014}). These drifts plus thruster corrections can produce abrupt changes in the measured flux that must be corrected or ameliorated to obtain photometric noise levels of the order of $2-4$ times greater than those for the original {\it Kepler} mission, where pointing with three reaction wheels was stable to a small fraction of a pixel. 

\citet{vandenburg2014}, \citet{armstrong2015} and \citet{lund2015} have devised pipelines to create improved pixel masks and/or to correct for pointing drift for each star in the K2 campaign fields. These pipelines are not equivalent, and the results are dependent on the variability in the stars. RR\,Lyr stars, for example, typically have periods of the order of 12\,h (less than 0.5\,d for RRc stars and greater than 0.5\,d for RRab stars), with rapid increases in brightness followed by a slower decrease (RRab stars), or a more sinusoidal variation in the case of RRc stars. The amplitudes are relatively large -- several tenths of a magnitude to well over 1\,mag -- so that the rise and fall times of the light variation are similar to the 5.9-h thruster firing schedule of the K2 mission. This can lead to overcorrection of the instrumental slopes of the light curves of RR\,Lyr stars, so care must be taken when using pipeline data for these stars, and when creating custom data rectification, either automatically or by hand. 

For EPIC\,201585823 we have examined and compared the pipeline reductions and show the results in Section\,\ref{sec:pipelines} below. A consequence of the large amplitude of EPIC\,201585823 and its pulsation timescale resonating with the thruster firing times leads to pipeline-corrected light curves that sometimes are more non-sinusoidal, hence with more combination frequency peaks, than in `raw' data that have been extracted with a customised mask and corrected for flat field and cosmic ray exclusion, but not corrected for changes in the satellite pointing. Thus, for the astrophysical analysis of this star we have chosen to use `raw' data extracted by us as part of the K2P$^2$ pipeline (K2-Pixel-Photometry; \citealt{lund2015}).\footnote{Available from the {\it Kepler} Asteroseismic Science Operations Centre; \newline http://kasoc.phys.au.dk} This may be the preferred choice for analysis of RR\,Lyr stars from K2 data. The K2P$^2$ pipeline automatically creates a mask that encompasses all pixels with significant signal for a star by using a summed image for mask selection. This therefore encompasses the approximately 1 pixel drift between the 5.9-hr thruster firings. The mask we used is shown in comparison with those of other pipelines in Section\,\ref{sec:pipelines}. 

\subsection{Frequency analysis of the K2P$^2$ SC raw light curve} 

\begin{figure*}
\centering	
\includegraphics[width=0.99\linewidth,angle=0]{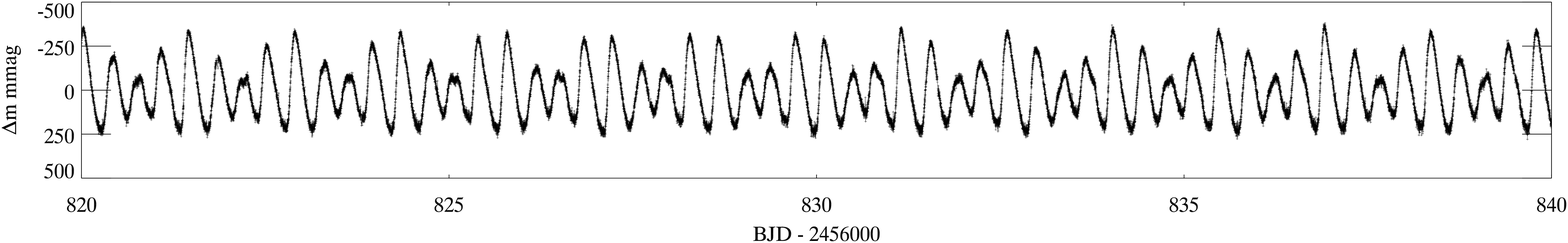}	
\caption{A section of the K2P$^2$ SC raw light curve of EPIC\,201585823 spanning 20\,d showing the high amplitude light variations. Almost all of the variation in the light curve is explained by the nonlinear interaction of  the fundamental and first-overtone radial p~modes in this RRd star.}
\label{fig:20158_lc}
\end{figure*}

The K2P$^2$ `raw' short cadence (SC; 58.9\,s integrations) data were analysed for their component frequencies. These data were obtained using a pipeline-generated mask encompassing all pixels with significant signal for the star, were flat-fielded, and had outliers removed, particularly at the times of thruster firings. They were not corrected for instrumental variations caused by pointing drift and re-setting, for reasons given in Section\,\ref{sec:data} above. The data set consists of 112\,545 points spanning 80.1\,d. Fig.\,\ref{fig:20158_lc} shows a typical 20-d section of the light curve where the variations, dominated by the fundamental and first-overtone radial modes, can be seen. Note that the light curve is nonlinear, with larger `upward' than `downward' excursions about the mean. We explain that upward light curve shape in Section\,\ref{sec:phasors} below, following \citet{kurtzetal2015}.

We performed a frequency analysis on the EPIC\,201585823 K2 C1 K2P$^2$ raw SC data set. We used the interactive light curve and amplitude spectrum tools in the programme {\small PERIOD04} \citep{lenz&breger2004}. We then used a Discrete Fourier Transform \citep{kurtz85} and our own least-squares and nonlinear least-squares fitting programs to find the frequencies, amplitudes and phases to describe the light curves. After setting the mean of the data set to zero, we fitted a cosine function, $\Delta m = A \cos (2 \pi f (t - t_0) + \phi)$, for each frequency in the data in magnitudes, thus defining our convention for the phases in this paper. The zero point of the time scale for the phases is ${\rm BJD}~{2\,456\,850.30000}$. Our routines and {\small PERIOD04} are in agreement. 

\subsection{The fundamental and first-overtone radial modes}

\begin{figure}
\centering	
\includegraphics[width=1.0\linewidth,angle=0]{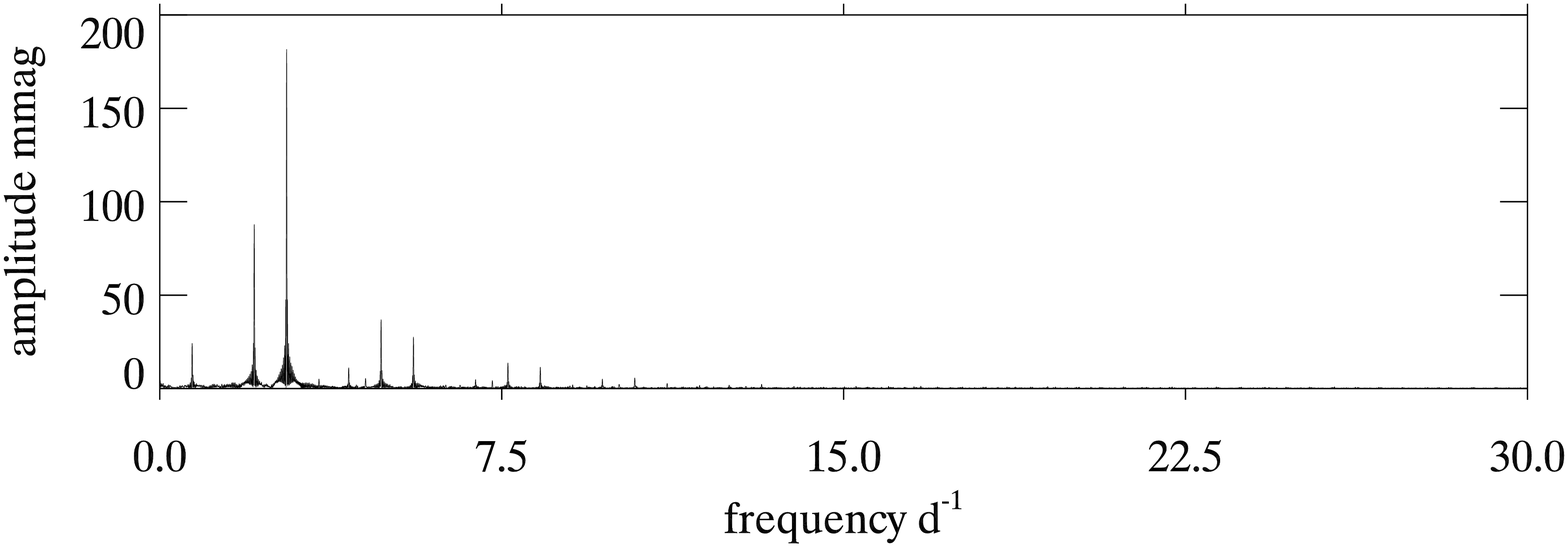}	
\includegraphics[width=1.0\linewidth,angle=0]{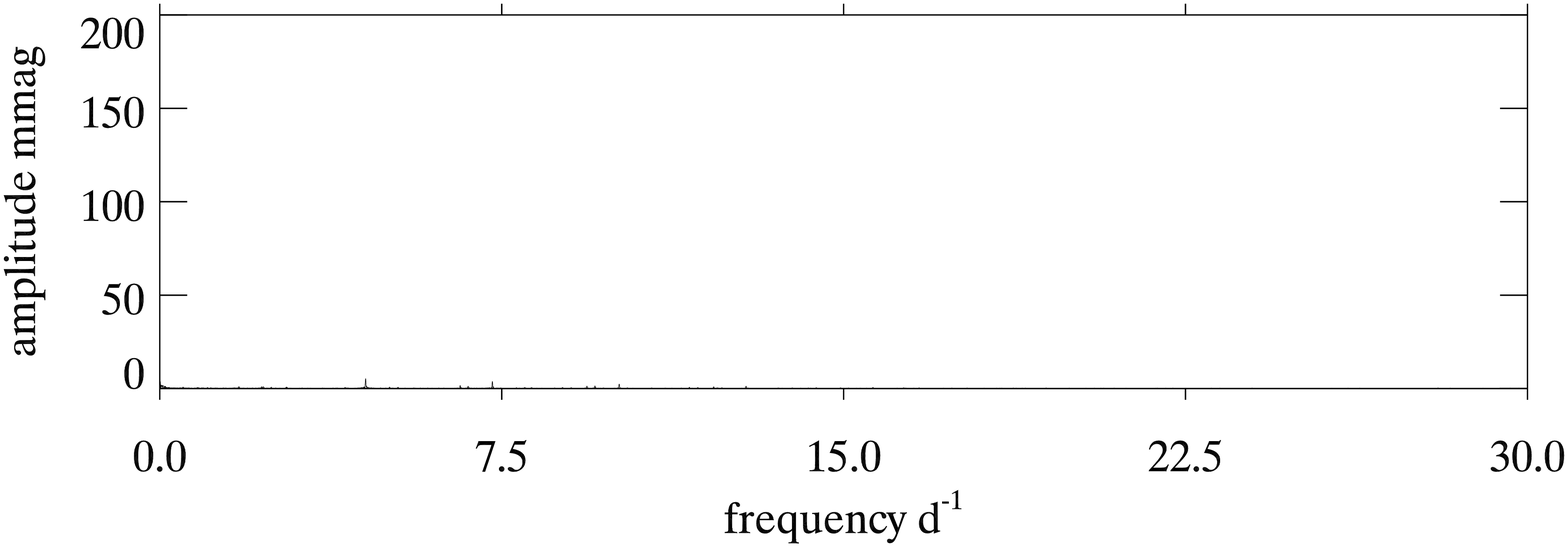}	
\caption{Top: An amplitude spectrum of the {\it Kepler} K2 C1 raw SC data for EPIC\,201585823 out to 30\,d$^{-1}$.  The two highest amplitude peaks, $\nu_1$ and $\nu_2$, are generated by the principal pulsation modes, which we identify as the fundamental and first-overtone radial modes based on their period ratio of 0.745. Most of the other peaks seen in this panel are combination frequencies. 
Bottom panel: After pre-whitening by $\nu_1$, $\nu_2$ and 42 combination frequencies up to order $5\nu$ and with amplitudes greater than 0.15\,mmag. This plot has been kept on the same ordinate scale as the top panel to make clear that most of the variance is in the two nonlinearly interacting radial modes. There is significant variance left that is shown below with this panel at a higher ordinate scale in Fig.\,\ref{fig:20158_ft_2}}
\label{fig:20158_ft_1}
\end{figure}

The light variations of EPIC\,201585823 are dominated by two peaks in the amplitude spectrum, which we identify as the fundamental and first-overtone radial modes from their period ratio, and that can be seen in the top panel of Fig.\,\ref{fig:20158_ft_1}. We identified the two highest peaks and fitted them by linear and nonlinear least-squares to the data with the results shown in Table\,\ref{tab:20158_1}. For generality, we label the fundamental radial mode frequency $\nu_1$, and the first-overtone radial mode frequency $\nu_2$. Traditionally in studies of RR\,Lyr stars these are called $f_0$ and $f_1$, respectively, or `F' and `1O', for the fundamental and first-overtone radial modes.  The period ratio, $0.744770 \pm 0.000003$, is consistent with other RRd stars, and with RR\,Lyr star models for fundamental and first-overtone radial pulsation. The error in amplitude is estimated to be $1\sigma = 0.07$\,mmag from the fit of the two base frequencies and their combination terms as seen in Table\,\ref{tab:20158_2}. The frequency errors and phase errors are proportional to the amplitude error \citep{montgomery-odonoghue99}, hence are  scaled appropriately.

Pulsations in RR\,Lyr stars, including RRd stars, are known to be non-sinusoidal, with many harmonics and combination frequencies  (see, e.g., \citealt{jerz1977}; \citealt{chadid2012}). We therefore searched for combination frequencies for $\nu_1$ and $\nu_2$ up to terms of order $5\nu$. While combination frequency peaks with terms higher than $5\nu$ are identifiable in K2 data for this star at amplitudes higher than our 0.15\,mmag cutoff, we chose not to include higher-order terms to keep the number of combination frequencies low for convenience of visualisation, and to avoid chance coincidences of combination frequencies with each other, and with peaks near the noise level. This does mean that some combination frequency peaks with amplitudes greater than 0.15\,mmag remain in our amplitude spectra. 

We did not search the amplitude spectrum for significant peaks that were then identified with combination frequencies. Instead, we hypothesised a combination frequency model and calculated the exact values of the combination frequencies from the base frequencies, $\nu_1$ and $\nu_2$, given in Table\,\ref{tab:20158_1}; we then fitted the base frequencies and their combinations to the data by linear least-squares. This generated 42 combination frequencies, in addition to the two base frequencies, with amplitudes greater than or equal to 0.15\,mmag ($2\sigma$ of our amplitude error); we discarded a further 16 combination frequencies with amplitudes lower than this limit. We then fitted the adopted 44 frequencies by linear least-squares to the data; the results are listed in Table\,\ref{tab:20158_2}. 

Pre-whitening the data by the 44 frequencies given in Table\,\ref{tab:20158_2} results in the amplitude spectrum shown in the bottom panel of Fig.\,\ref{fig:20158_ft_1}. This has intentionally been kept on the same ordinate scale as the top panel of the figure for impact: Most of the variance is in the fundamental and first-overtone radial mode frequencies, and their combination frequencies, as is typical of RRd stars. However, with a higher ordinate scale there are many real peaks still to be examined, as we discuss in detail next in Section\,\ref{sec:additional} and show in Fig.\,\ref{fig:20158_ft_2}. 

\begin{table}
\centering
\caption[]{A nonlinear least-squares fit of only the two principal pulsation mode frequencies of EPIC\,201585823, identified with the radial fundamental ($f_0$) and first-overtone ($f_1$) modes.   The zero point of the time scale for the phases is ${\rm BJD}~{2\,456\,850.30000}$.}
\begin{tabular}{lrrr}
\toprule
\multicolumn{1}{c}{labels} &
\multicolumn{1}{c}{frequency} & \multicolumn{1}{c}{amplitude} &
\multicolumn{1}{c}{phase} \\
&\multicolumn{1}{c}{d$^{-1}$} & \multicolumn{1}{c}{mmag} &
\multicolumn{1}{c}{radians}  \\
& & \multicolumn{1}{r}{$\pm 0.07$} &\\
\toprule

$ \nu_1 \,(f_0)$ & $  2.072143 \pm 0.000005$  & $  88.34     $ & $  1.1525   \pm  0.0008$  \\
$ \nu_2 \,(f_1)$ & $  2.782259 \pm 0.000003$  & $  181.70      $ & $  -2.1659   \pm  0.0004$  \\

\bottomrule

\end{tabular}
\label{tab:20158_1}
\end{table}

\begin{table}
\centering
\caption[]{A linear least-squares fit of the two principal pulsation mode frequencies of EPIC\,201585823, identified with the radial fundamental and first-overtone modes, plus 42 combination frequencies up to order $5\nu$ and with amplitudes greater than 0.15\,mmag.   The zero point of the time scale for the phases is ${\rm BJD}~{2\,456\,850.30000}$.}
\begin{tabular}{crrr}
\toprule
\multicolumn{1}{c}{labels} &
\multicolumn{1}{c}{frequency} & \multicolumn{1}{c}{amplitude} &
\multicolumn{1}{c}{phase} \\
&\multicolumn{1}{c}{d$^{-1}$} & \multicolumn{1}{c}{mmag} &
\multicolumn{1}{c}{radians}  \\
& & \multicolumn{1}{r}{$\pm 0.07$} &\\
\toprule

$-4\nu_1   +   3\nu_2     $ & $   0.058205   $ & $   1.45   $ & $   -0.8924  \pm   0.0498 $ \\ 
 $3\nu_1   -   2\nu_2     $ & $   0.651911   $ & $   0.30   $ & $   -0.7876  \pm   0.2441 $ \\ 
 $-\nu_1   +   \nu_2     $ & $   0.710116   $ & $   24.09   $ & $   0.6326  \pm   0.0031 $ \\ 
 $-5\nu_1   +   4\nu_2     $ & $   0.768321   $ & $   0.38   $ & $   -3.0419  \pm   0.1966 $ \\ 
 $2\nu_1   -   \nu_2     $ & $   1.362027   $ & $   1.75   $ & $   -0.2407  \pm   0.0421 $ \\ 
 $-2\nu_1   +   2\nu_2     $ & $   1.420232   $ & $   1.80   $ & $   -0.1891  \pm   0.0409 $ \\ 
 $5\nu_1   -   3\nu_2     $ & $   2.013938   $ & $   0.15   $ & $   -2.7546  \pm   0.4799 $ \\ 
 $ \nu_1        $ & $   2.072143   $ & $   88.43   $ & $   1.1525  \pm   0.0008 $ \\ 
 $-3\nu_1   +   3\nu_2     $ & $   2.130348   $ & $   0.31   $ & $   0.3009  \pm   0.2353 $ \\ 
 $  \nu_2     $ & $   2.782259   $ & $   181.64   $ & $   -2.1657  \pm   0.0004 $ \\ 
 $3\nu_1   -   \nu_2     $ & $   3.434170   $ & $   0.39   $ & $   -0.3131  \pm   0.1890 $ \\ 
 $-\nu_1   +   2\nu_2     $ & $   3.492375   $ & $   4.68   $ & $   2.9487  \pm   0.0158 $ \\ 
 $-5\nu_1   +   5\nu_2     $ & $   3.550580   $ & $   0.21   $ & $   -2.4231  \pm   0.3502 $ \\ 
 $2\nu_1        $ & $   4.144286   $ & $   11.14   $ & $   -0.0980  \pm   0.0066 $ \\ 
 $-2\nu_1   +   3\nu_2     $ & $   4.202491   $ & $   0.93   $ & $   0.6101  \pm   0.0792 $ \\ 
 $\nu_1   +   \nu_2     $ & $   4.854402   $ & $   36.89   $ & $   3.1205  \pm   0.0020 $ \\ 
 $  2\nu_2     $ & $   5.564518   $ & $   27.85   $ & $   0.5041  \pm   0.0026 $ \\ 
 $3\nu_1        $ & $   6.216429   $ & $   1.29   $ & $   -0.6625  \pm   0.0572 $ \\ 
 $-\nu_1   +   3\nu_2     $ & $   6.274634   $ & $   2.05   $ & $   2.2999  \pm   0.0359 $ \\ 
 $2\nu_1   +   \nu_2     $ & $   6.926545   $ & $   4.86   $ & $   2.2648  \pm   0.0152 $ \\ 
 $-2\nu_1   +   4\nu_2     $ & $   6.984750   $ & $   0.84   $ & $   2.1840  \pm   0.0877 $ \\ 
 $\nu_1   +   2\nu_2     $ & $   7.636661   $ & $   13.81   $ & $   -1.0229  \pm   0.0053 $ \\ 
 $-3\nu_1   +   5\nu_2     $ & $   7.694866   $ & $   0.21   $ & $   0.8220  \pm   0.3556 $ \\ 
 $  3\nu_2     $ & $   8.346777   $ & $   11.76   $ & $   2.9695  \pm   0.0063 $ \\ 
 $3\nu_1   +   \nu_2     $ & $   8.998688   $ & $   1.13   $ & $   0.8866  \pm   0.0653 $ \\ 
 $-\nu_1   +   4\nu_2     $ & $   9.056893   $ & $   1.86   $ & $   -2.4803  \pm   0.0397 $ \\ 
 $2\nu_1   +   2\nu_2     $ & $   9.708804   $ & $   5.08   $ & $   -1.5072  \pm   0.0145 $ \\ 
 $-2\nu_1   +   5\nu_2     $ & $   9.767009   $ & $   0.65   $ & $   -2.1820  \pm   0.1136 $ \\ 
 $\nu_1   +   3\nu_2     $ & $   10.418920   $ & $   5.57   $ & $   2.4194  \pm   0.0132 $ \\ 
 $4\nu_1   +   \nu_2     $ & $   11.070831   $ & $   0.30   $ & $   1.5387  \pm   0.2462 $ \\ 
 $  4\nu_2     $ & $   11.129036   $ & $   2.86   $ & $   -0.3051  \pm   0.0258 $ \\ 
 $3\nu_1   +   2\nu_2     $ & $   11.780947   $ & $   0.72   $ & $   -1.6379  \pm   0.1027 $ \\ 
 $-\nu_1   +   5\nu_2     $ & $   11.839152   $ & $   1.27   $ & $   0.1206  \pm   0.0580 $ \\ 
 $2\nu_1   +   3\nu_2     $ & $   12.491063   $ & $   1.51   $ & $   1.5338  \pm   0.0488 $ \\ 
 $\nu_1   +   4\nu_2     $ & $   13.201179   $ & $   2.18   $ & $   -0.9836  \pm   0.0339 $ \\ 
 $  5\nu_2     $ & $   13.911295   $ & $   1.26   $ & $   2.9376  \pm   0.0583 $ \\ 
 $3\nu_1   +   3\nu_2     $ & $   14.563206   $ & $   0.21   $ & $   0.2337  \pm   0.3578 $ \\ 
 $2\nu_1   +   4\nu_2     $ & $   15.273322   $ & $   0.77   $ & $   -2.4566  \pm   0.0953 $ \\ 
 $5\nu_1   +   2\nu_2     $ & $   15.925233   $ & $   0.15   $ & $   -0.0189  \pm   0.4893 $ \\ 
 $\nu_1   +   5\nu_2     $ & $   15.983438   $ & $   1.11   $ & $   1.7042  \pm   0.0663 $ \\ 
 $3\nu_1   +   4\nu_2     $ & $   17.345465   $ & $   0.41   $ & $   3.1159  \pm   0.1795 $ \\ 
 $2\nu_1   +   5\nu_2     $ & $   18.055581   $ & $   0.51   $ & $   0.5658  \pm   0.1439 $ \\ 
 $5\nu_1   +   3\nu_2     $ & $   18.707492   $ & $   0.20   $ & $   -1.9411  \pm   0.3752 $ \\ 
 $3\nu_1   +   5\nu_2     $ & $   20.127724   $ & $   0.21   $ & $   -0.1201  \pm   0.3580 $ \\ 
 
\bottomrule

\end{tabular}
\label{tab:20158_2}
\end{table}

\subsection{A third mode frequency - a nonradial mode }
\label{sec:additional}

\begin{figure}
\centering	
\includegraphics[width=1.0\linewidth,angle=0]{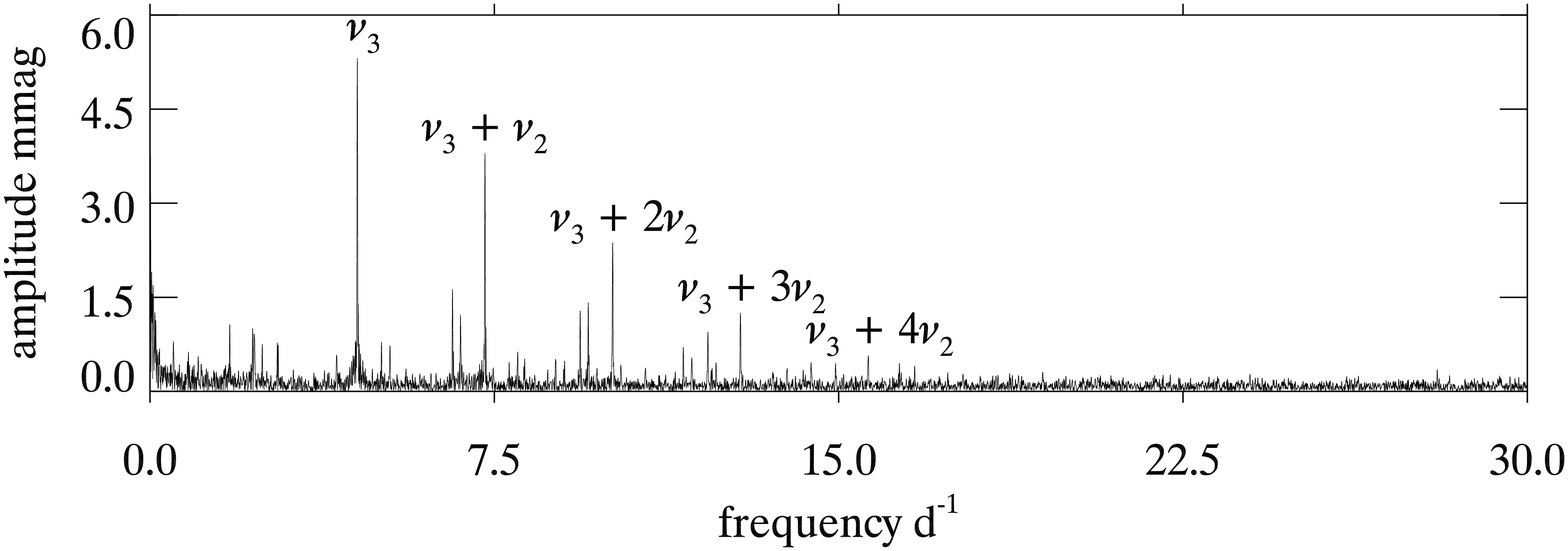}	
\includegraphics[width=1.0\linewidth,angle=0]{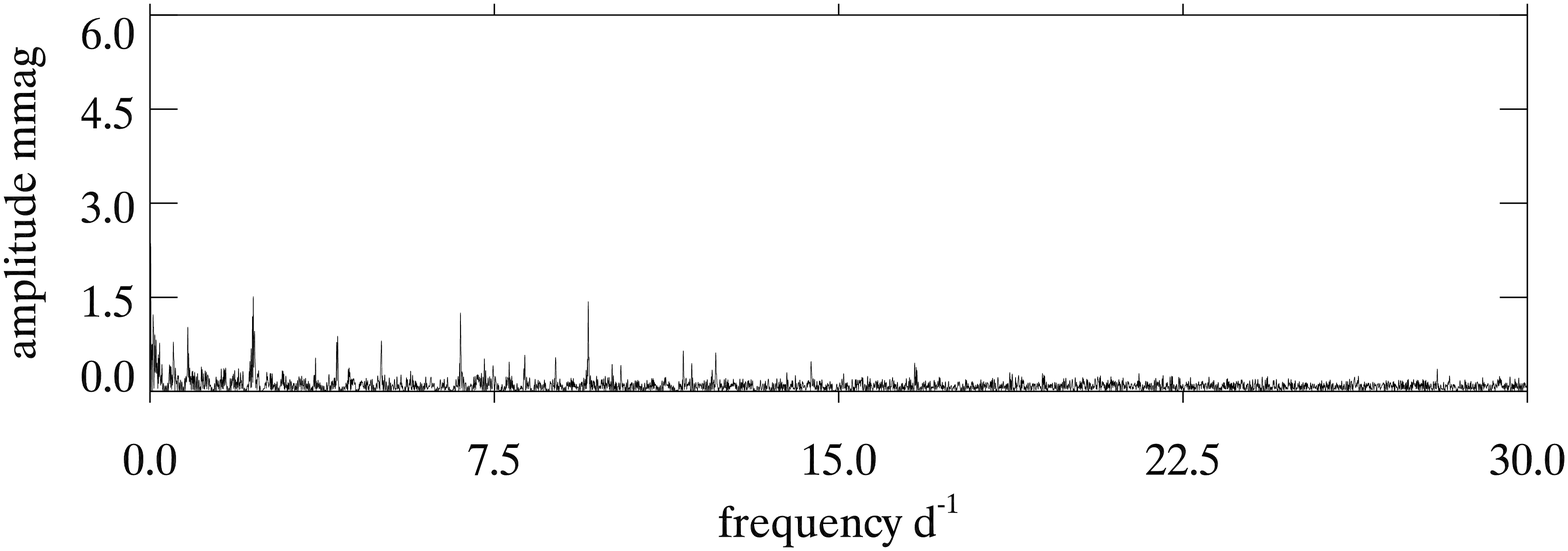}	
\includegraphics[width=1.0\linewidth,angle=0]{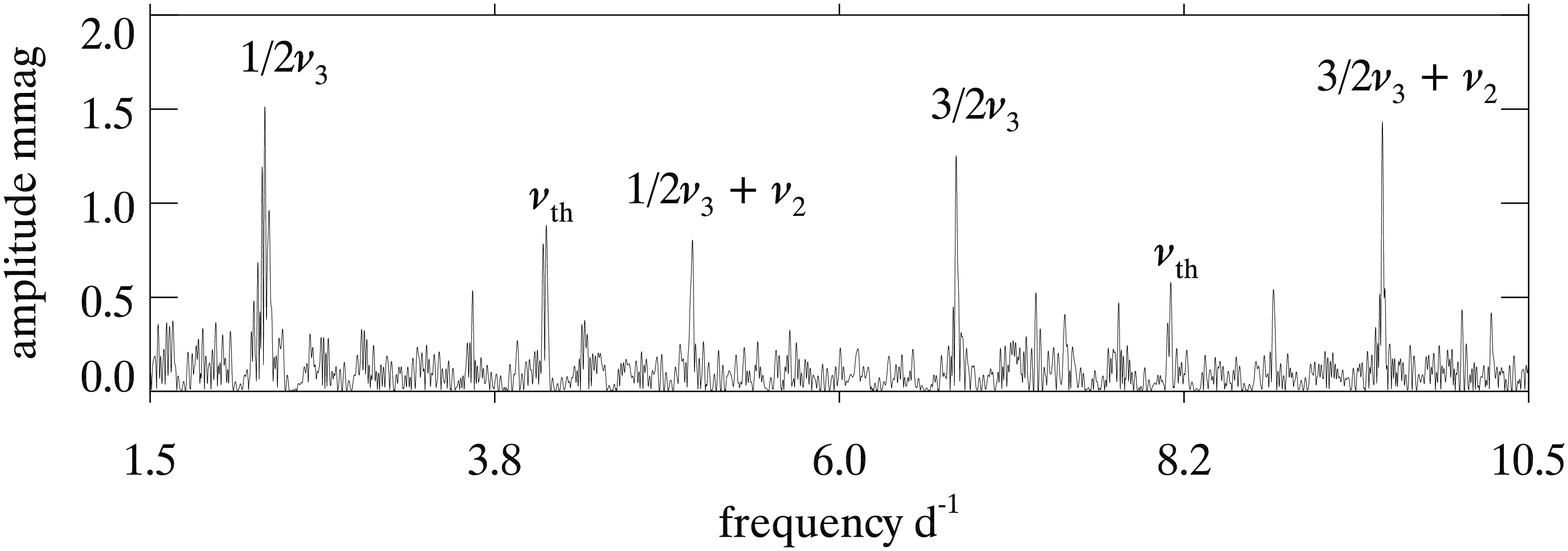}	
\caption{Top: The same amplitude spectrum as in the bottom panel of Fig.\,\ref{fig:20158_ft_1} after pre-whitening by $\nu_1$, $\nu_2$ and 42 combination frequencies up to order $5\nu$ and with amplitudes greater than 0.15\,mmag, but on a larger scale. The highest peak is $\nu_3 =4.51456$\,d$^{-1}$. Four combination frequencies with $\nu_2$ are labelled, indicating that $\nu_3$ is a mode frequency.  Middle: The amplitude spectrum of the residuals to a 414-frequency fit with $\nu_1$, $\nu_2$ and $\nu_3$, plus combination frequencies up to order $5\nu$ and with amplitudes greater than or equal to 0.15\,mmag. Bottom: An expanded look at the middle panel showing the subharmonics of $\nu_3$. For clarity, combination peaks of the subharmonics with $\nu_2$ with relatively high amplitude are marked. Two frequencies are marked as $\nu_{\rm th}$; these are artefacts at the thruster frequency, 4.08\,d$^{-1}$, and its harmonic, 8.16\,d$^{-1}$. Multiple peaks are seen for those, since the thruster firings are not purely periodic. These artefacts, and higher harmonics of the thruster firing frequency are common in K2 data sets.}
\label{fig:20158_ft_2}
\end{figure}

After pre-whitening by $\nu_1$, $\nu_2$ and their 42 combination terms listed in Table\,\ref{tab:20158_2} we find a third frequency, $\nu_3 = 4.51456 \pm 0.00008$\,d$^{-1}$, which has a number of combination frequencies with $\nu_1$ and $\nu_2$, many of which are evident in the top panel of Fig.\,\ref{fig:20158_ft_2} (which is the same as the bottom panel of Fig.\,\ref{fig:20158_ft_1}, but with higher ordinate scale). The four highest amplitude combination frequencies with $\nu_2$ are marked, and others can be seen. That $\nu_3$ couples to $\nu_1$ and $\nu_2$ argues that it is a mode frequency. It does not coincide with any combination frequency of $\nu_1$ and $\nu_2$.

A nonlinear least-squares fit of $\nu_1$, $\nu_2$ and $\nu_3$ is given in Table\,\ref{tab:20158_3} (but not their 411 combination frequencies with amplitudes greater than or equal to 0.15\,mmag, which are too numerous to list). From this we associate $\nu_3$ with a nonradial pulsation mode, based on its period ratio with the highest amplitude mode, the first-overtone radial mode at $\nu_2$; this ratio is  $0.616285 \pm 0.000012$. 

\begin{table}
\centering
\caption[]{A nonlinear least-squares fit of the two principal pulsation mode frequencies of EPIC\,201585823, identified with the radial fundamental ($f_0$) and first-overtone ($f_1$) modes, and the third mode frequency $\nu_3$, which arises from a nonradial mode.  The zero point of the time scale for the phases is ${\rm BJD}~{2\,456\,850.30000}$.}
\begin{tabular}{lrrr}
\toprule
\multicolumn{1}{c}{labels} &
\multicolumn{1}{c}{frequency} & \multicolumn{1}{c}{amplitude} &
\multicolumn{1}{c}{phase} \\
&\multicolumn{1}{c}{d$^{-1}$} & \multicolumn{1}{c}{mmag} &
\multicolumn{1}{c}{radians}  \\
& & \multicolumn{1}{r}{$\pm 0.07$} &\\
\toprule
$ \nu_1\,(f_0)$ & $  2.072142 \pm 0.000005$  & $  88.34     $ & $  1.1525   \pm  0.0008$  \\
$ \nu_2\,(f_1)$ & $  2.782258 \pm 0.000003$  & $  181.70      $ & $  -2.1659   \pm  0.0004$  \\
$ \nu_3 $ & $  4.514560 \pm 0.000080$  & $  5.60      $ & $  -0.5938   \pm  0.0125$  \\
\bottomrule

\end{tabular}
\label{tab:20158_3}
\end{table}

This is similar to the period ratios that were found in 11 other triple-mode RRd stars, which are listed in Table\,\ref{tab:rrtriples}, and in 260 double mode RRc stars that pulsate in the first-overtone radial mode and an addition nonradial mode (for lists of those stars, see: \citealt[b]{netzel2015a}; \citealt{moskalik2015}; \citealt{jurcsik2015}). We thus conclude that EPIC\,201585823 is a new member of a (currently) rare class of triple-mode RR\,Lyr stars. 

\begin{table*}
\centering
\caption[]{A list of the 12 known triple-mode RR Lyrae stars with period ratios near 0.61. The stars are ordered by increasing $P_2$, the period for the dominant first-overtone radial mode. $P_3/P_2$ is the ratio of the nonradial mode period ($P_3$) to that of the radial first overtone. All stars are remarkably similar in their period ratios.}
\begin{tabular}{lrrrrl}
\toprule

\multicolumn{1}{c}{name} &
\multicolumn{1}{c}{$P_1$} & \multicolumn{1}{c}{$P_2$} & \multicolumn{1}{c}{$P_3$} &\multicolumn{1}{c}{$P_3/P_2$} &
\multicolumn{1}{c}{reference} \\
& \multicolumn{1}{c}{d} &\multicolumn{1}{c}{d} & \multicolumn{1}{c}{d} & & \\
\toprule

OGLE-BLG-RRLYR-02027	 &  0.3799 & 0.2786 & 0.1702 &   0.6107 &   \citet{netzel2015a}\\
OGLE-BLG-RRLYR-07393  $^{\rm a}$ &  0.4627 & 0.3449 & 0.2126 &   0.6163 &   \citet{smolecetal2015a}  \\
V125 in M3                                           &  0.4709 & 0.3498 & 0.2158 &   0.6168 &   \citet{jurcsik2015}  \\
V13 in M3 $^{\rm a,b}$                         &  0.4795 & 0.3507 & 0.2153 &   0.6137 &   \citet{jurcsik2015}  \\
V68 in M3 &  0.4785                             & 0.3560 & 0.2187 &   0.6145 &   \citet{jurcsik2015}  \\
V87 in M3 &  0.4802                             & 0.3575 & 0.2208 &   0.6177 &   \citet{jurcsik2015}  \\
EPIC\,201585823                                   &  0.4826 & 0.3594 & 0.2215 &   0.6163 &   This paper  \\
CoRoT\,101368812                              &  0.4880 & 0.3636 & 0.2233 &   0.6141 &   \citet{chadid2012}  \\
EPIC\,60018653                                   &  0.5394 & 0.4023 & 0.2479 &   0.6162 &   \citet{molnar2015}  \\
AQ\,Leo                                                &  0.5498 & 0.4101 & 0.2547 &   0.6211 &   \citet{gruberbauer2007}  \\
EPIC\,60018662                                   &  0.5590 & 0.4175 & 0.2574 &   0.6166 &   \citet{molnar2015}  \\
OGLE-BLG-RRLYR-14031                   &  0.5761 & 0.4298 & 0.2647 &   0.6159 &   \citet{netzel2015a} \\
\bottomrule
\multicolumn{6}{l}{NOTES: a - Blazhko modulation of dominant radial modes; b - weak radial second overtone detected.} \\

\end{tabular}
\label{tab:rrtriples}
\end{table*}

\begin{table}
\centering
\caption[]{A comparison of the mode frequencies for EPIC\,201585823 and V87 in M3 \citep{jurcsik2015}. The stars are remarkably similar.}
\begin{tabular}{lrrrrc}
\toprule
&\multicolumn{2}{c}{EPIC\,201585823} & \multicolumn{2}{c}{V87 in M3} &\\
&\multicolumn{1}{c}{frequency} & \multicolumn{1}{c}{amplitude} &
\multicolumn{1}{c}{frequency} & \multicolumn{1}{c}{amplitude} & 
\multicolumn{1}{c}{frequency}\\
&\multicolumn{1}{c}{d$^{-1}$} & \multicolumn{1}{c}{mmag} &
\multicolumn{1}{c}{d$^{-1}$} & \multicolumn{1}{c}{mmag} &
\multicolumn{1}{c}{ratio}\\
\toprule

$ \nu_1$ & $  2.072142$  & $  88.3     $ &   $2.08260$ & $110.0$ & 1.005\\
$ \nu_2$ & $  2.782258$  & $  181.7      $ & $2.79728$ & $230.0$& 1.005\\
$ \nu_3 $ & $  4.514560$  & $  5.6      $ &   $4.52880$ & $7.8$ & 1.003\\
\bottomrule
\end{tabular}
\label{tab:comp}
\end{table}

To show the similarity of these triple-mode RR\,Lyr stars more specifically, in Table\,\ref{tab:comp} we compare the three mode frequencies of EPIC\,201585823 with those of V87 in M3 \citep{jurcsik2015}. The frequencies are very similar in these two stars, as they are also similar to those of the triple-mode RR Lyr variables V68 in M3 \citep{jurcsik2015} and CoRoT\,101368812 \citep{chadid2012}. These four triple-mode pulsators seem to be of similar stellar structure, hence in similar evolutionary states. In the other known stars of this type, frequencies of the modes are different, yet their ratios are still almost the same. In particular, the period ratio of the nonradial mode to the first-overtone radial mode is always in a narrow range of $0.611-0.621$. It will be interesting to see whether other triple-mode RR\,Lyr stars that are found in the future will also show such similarity. Many RR\,Lyr stars are being observed in the K2 campaign fields\footnote{http://keplergo.github.io/KeplerScienceWebsite/k2-approved-programs.html lists approved programmes for K2}, and it is reasonable to expect the discovery of more triple-mode RR Lyr variables.
 
After fitting $\nu_1$, $\nu_2$,  $\nu_3$ and combination frequencies to the data and pre-whitening, the middle and bottom panels of Fig.\,\ref{fig:20158_ft_2} show the amplitude spectrum of the residuals. Six peaks are marked: Two are very close to the exact subharmonics of $\nu_3$ at $\frac{1}{2}\nu_3$ and $\frac{3}{2}\nu_3$; two are combination frequencies between the subharmonics and $\nu_2$ at $\frac{1}{2}\nu_3 + \nu_2$ and $\frac{3}{2}\nu_3 + \nu_2$; and two, marked as $\nu_{\rm th}$, are artefacts at the thruster firing frequency and its harmonic. 

In Table 8 of \citet{moskalik2015} there are 13 stars (4 RRd and 9 RRc) with space photometry. Subharmonics were detected in all 4 RRd stars and in 6 of the 9 RRc stars. Thus, with space photometry these subharmonics are found in most RRc and RRd stars, and their properties are found to be the same in RRc and in RRd stars. These subharmonics can also be detected in ground-based studies. In the OGLE-IV Galactic Bulge survey data, \citet{netzel2015b} found 131 RRc stars with period ratios near 0.61. Of those, 26 showed a subharmonic at $\frac{1}{2} \nu$ of the nonradial mode frequency, and two more showed a subharmonic at $\frac{3}{2} \nu$.

Subharmonics are indicative of period doubling, and in EPIC\,201585823 they show unresolved, broad frequency peaks, which have also been seen in other stars showing the $\nu_3$ mode frequency; this shows that the period doubling changes are not constant in time \citep{moskalik2015}. We examine the frequency and/or amplitude changes of $\nu_3$ further in Section\,\ref{sec:subharmonics} below. The values of the subharmonics, $\frac{1}{2}\nu_3$ and $\frac{3}{2}\nu_3$, are not well-determined from the full data set. Their frequency and amplitude variability give rise to asymmetric and multiple peaks in the amplitude spectrum, as can be seen in the bottom panel of Fig.\,\ref{fig:20158_ft_2}. Selecting the highest peak in both cases gives formal values of $\frac{1}{2}\nu_3 = 2.2492 \pm 0.0003$\,d$^{-1}$ and $\frac{3}{2}\nu_3 = 6.7625 \pm 0.0004$\,d$^{-1}$; these are nominally just resolved from the exact half integer values of 2.2573\,d$^{-1}$ and 6.7718\,d$^{-1}$, respectively. 

While there are further significant peaks seen in the bottom panel of Fig.\,\ref{fig:20158_ft_2}, we chose to terminate the frequency analysis at this point, leaving the question of whether these other peaks belong to other pulsation modes, combination frequencies, instrumental artefacts or other variations in the data. Our caution arises from the difficulty of rectifying fully the K2 data for RR\,Lyr stars. We illustrate some of the problems with that in Section\,\ref{sec:pipelines} below.

\subsection{Frequency and amplitude variability}
\label{sec:subharmonics}

\begin{figure}
\centering	
\includegraphics[width=1\linewidth,angle=0]{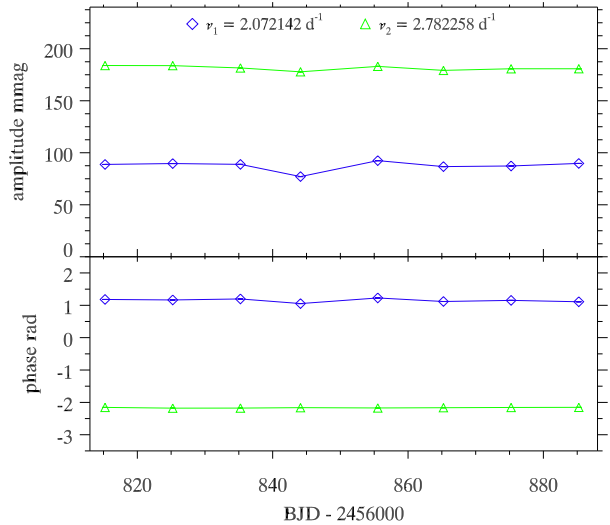}	
\caption{Tracking plot showing optimised values of amplitude and phase at fixed frequency using linear least-squares in 10-d bins for the fundamental mode frequency $\nu_1$ and the radial first overtone mode frequency $\nu_2$. 1$\sigma$ error bars calculated from the least-squares fit are plotted, but are generally smaller than the data points. The zero point of the time scale for the phases is ${\rm BJD}~{2\,456\,850.30000}$.}
\label{fig:tracking 1}
\end{figure}

\begin{figure*}
\centering	
\includegraphics[width=0.49\linewidth,angle=0]{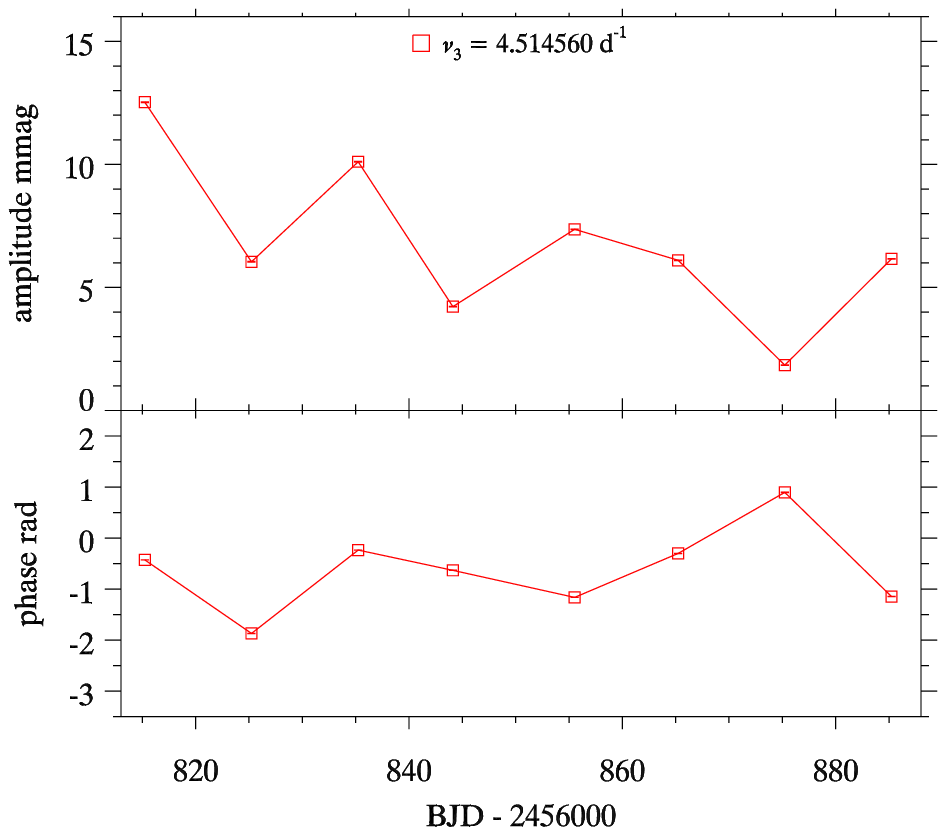}	
\includegraphics[width=0.49\linewidth,angle=0]{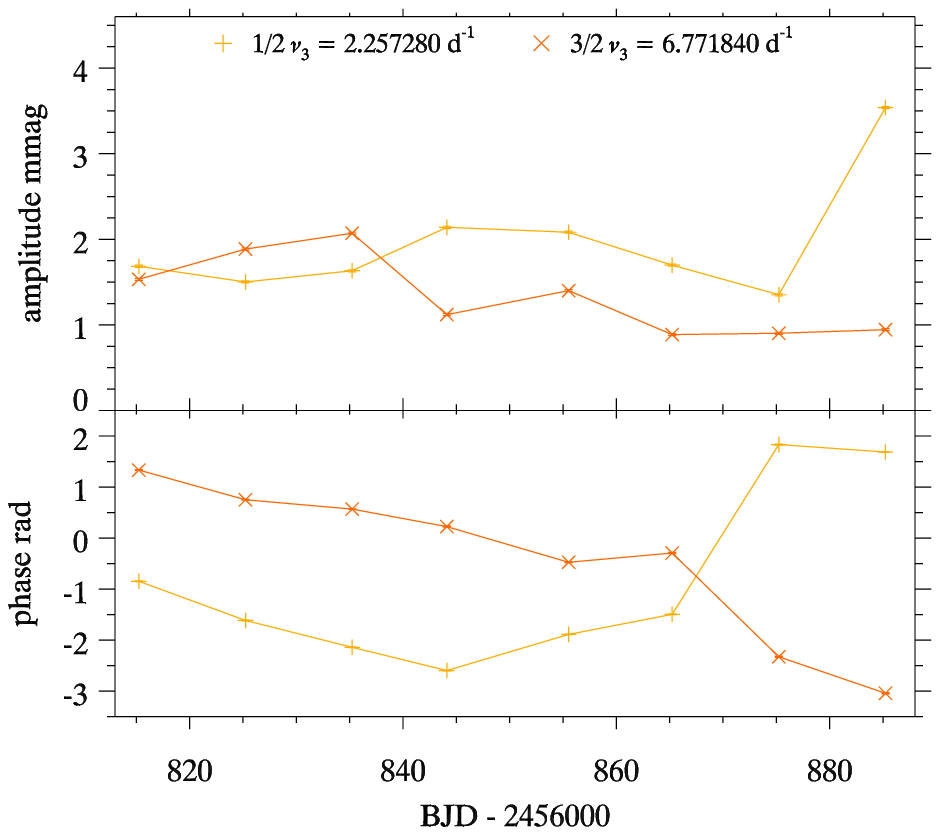}	
\caption{Tracking plots showing optimised values of amplitude and phase at fixed frequency using linear least-squares in 10-d bins. Left panel: The nonradial mode frequency $\nu_3$. Right panel: Subharmonics of $\nu_3$. 1$\sigma$ error bars are plotted as calculated from the least-squares fit; they are generally smaller than the data points. The zero point of the time scale for the phases is ${\rm BJD}~{2\,456\,850.30000}$.  Both the amplitudes and the phases (thus frequencies) are variable.}
\label{fig:tracking 2}
\end{figure*}

We employed the methodology of \citet{bowman2014}, who tracked amplitude and phase at fixed frequency through a series of discrete time bins to study variable pulsation amplitudes and frequencies in a $\delta$~Sct star, KIC\,7106205. For EPIC~201585823, we used the frequencies given in Table~\ref{tab:20158_3}, which were determined from the entire 80.1-d dataset. We then divided the dataset into eight bins, each 10~d in length. Values of amplitude and phase were optimised at fixed frequency using linear least-squares in each time bin and plotted against time. The results are shown in Fig.~\ref{fig:tracking 1} for $\nu_1$ and $\nu_2$, and in the left panel of Fig.~\ref{fig:tracking 2} for $\nu_3$. The zero point of the time scale for the phases is ${\rm BJD}~{2\,456\,850.30000}$. We chose to study the observed phase variation by keeping the frequency fixed in a linear least-squares fit, which is equivalent to studying frequency variation at fixed phase.

The radial fundamental  and first-overtone mode frequencies, $\nu_1$ and $\nu_2$, are shown in Fig.~\ref{fig:tracking 1}.  They exhibit little amplitude or phase variation over the 80.1~d of K2 observations. In the absence of the Blazhko effect, for which we have no clear evidence in this star, this result is not unexpected for high-amplitude, low-order radial p~modes. Non-Blazhko RR\,Lyr stars have stable frequencies and amplitudes over this time span; typical changes in frequency of radial modes are of order less than 0.01\,per\,cent (e.g. \citealt{jurcsik2015}). There is a small drop in amplitude in the middle of the observations (approximately ${\rm BJD}~2\,456\,845$) for both $\nu_1$ and $\nu_2$ in Fig.~\ref{fig:tracking 1}; this may be instrumental in origin since that bin contains a gap in the observations. 

The third frequency $\nu_3$ (shown in the left panel of Fig.~\ref{fig:tracking 2}) and its subharmonics $\frac{1}{2} \nu_3$ and $\frac{3}{2} \nu_3$ (shown in the right panel of Fig.~\ref{fig:tracking 2}), exhibit considerable amplitude and phase variation over 80.1~d. This behaviour is the same as that seen in RRc stars using similar techniques reported  by \citet[their figs 6 and 7]{moskalik2015}, \citet[their fig. 11]{szabo2014} and \citet[their fig. 5]{molnar2015}. We conclude that the nonradial mode in these stars commonly shows amplitude and frequency variability. 

\subsection{Phasor plots for the radial modes}
\label{sec:phasors}

\begin{figure}
\centering	
\includegraphics[width=0.6\linewidth,angle=0]{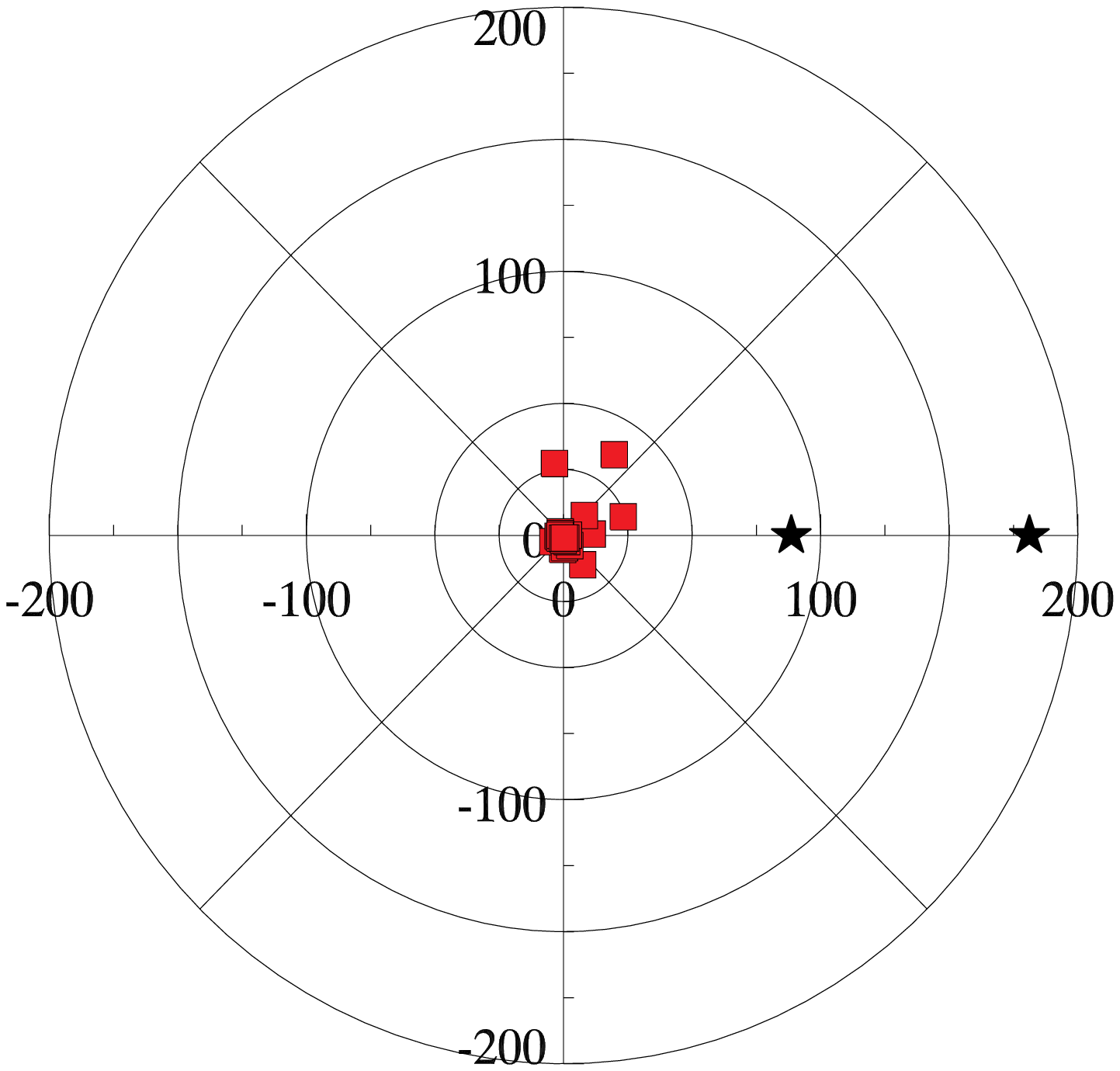}	
\includegraphics[width=0.6\linewidth,angle=0]{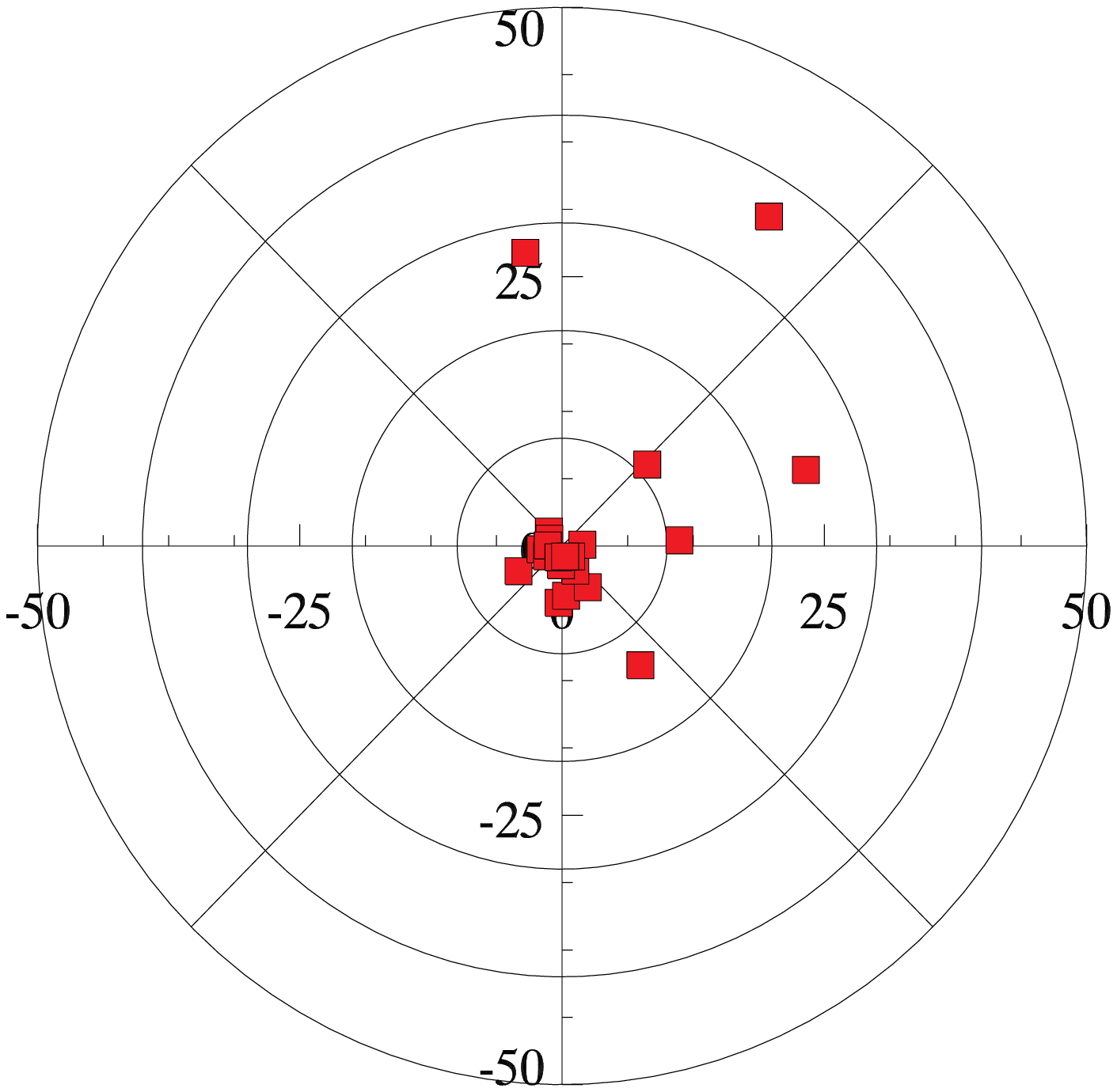}	
\caption{Phasor plots for EPIC\,201585823. These polar plots show amplitude in mmag as the radial coordinate and phase in radians as the angular coordinate. The relative phase is defined as $\phi_{\rm r} = \phi_{\rm obs} - \phi_{\rm calc} = \phi_{\rm obs} - (n\phi_i + m\phi_j)$. The convention is such that upward shaped light curves have combination frequency phases near to zero phase in the right-hand section of the diagrams \citep{kurtzetal2015}. The top panel shows the amplitudes of the base frequencies (black stars), $\nu_1$ and $\nu_2$, with their phases set to zero. Because of the large relative amplitudes of the base frequencies compared to the combination frequencies, the bottom plot shows those combination frequency amplitudes and phases at a larger scale. }
\label{fig:phasor}
\end{figure}

\citet{kurtzetal2015} showed that the shapes of the light curves of strongly nonlinearly pulsating Slowly Pulsating B (SPB) stars  and $\gamma$~Dor stars are governed by the phases of the combination frequencies. In particular, light curves with an `upward' shape, such as that seen for EPIC\,201585823 in Fig.\,\ref{fig:20158_lc}, are the result of combination frequencies having phases close to zero in comparison with the base frequencies. We show that is the case for EPIC\,201585823 in Fig.\,\ref{fig:phasor} where the highest amplitude combination frequencies of $\nu_1$ and $\nu_2$ (see Table\,\ref{tab:20158_2}) lie to the right with relative phases near zero, as expected for an upward light curve. 

The upward nonlinearity of the RR\,Lyr light curves has previously been considered to be normal and understood.  That the relative phases of the combination frequencies to those of the base mode frequencies in these stars are as predicted by  \citet{kurtzetal2015} supports their interpretation of upward and downward nonlinearities in pulsating star light curves in general. 

\section{A comparison of the different pipelines}
\label{sec:pipelines}

\begin{figure}
\centering	
\includegraphics[width=0.99\linewidth,angle=0]{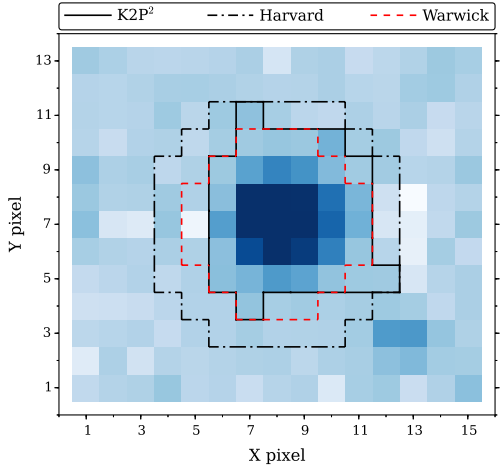}	
\caption{Pixel masks adopted for EPIC 20158523 for the different pipelines used in this work (see legend). The colours depict the summed image with the background subtracted from individual frames (see \citealt{lund2015}); the colour scale is logarithmic and goes from light (low flux) to dark blue (high flux). For the Harvard pipeline we show the mask obtained from the MAST data products. The faint target in the lower right part of the plot, centred on pixel (13,3), was also detected by the K2P$^2$ pipeline but we have omitted its mask in this plot.}
\label{fig:masks}
\end{figure}

The analysis of EPIC\,201585823 in this paper used K2P$^2$ SC `raw' data with a pipeline-created mask to capture all pixels with significant variability,  corrected for flat-field and removed obvious outliers. Fig.\,\ref{fig:masks} shows the mask we used in comparison with the masks used by \citet{vandenburg2014} and \citet{armstrong2015}. We did not correct for the pointing changes caused by drift and thruster firings, nor for possible other long-term trends. 

\begin{figure*}
\centering	
\includegraphics[width=0.49\linewidth,angle=0]{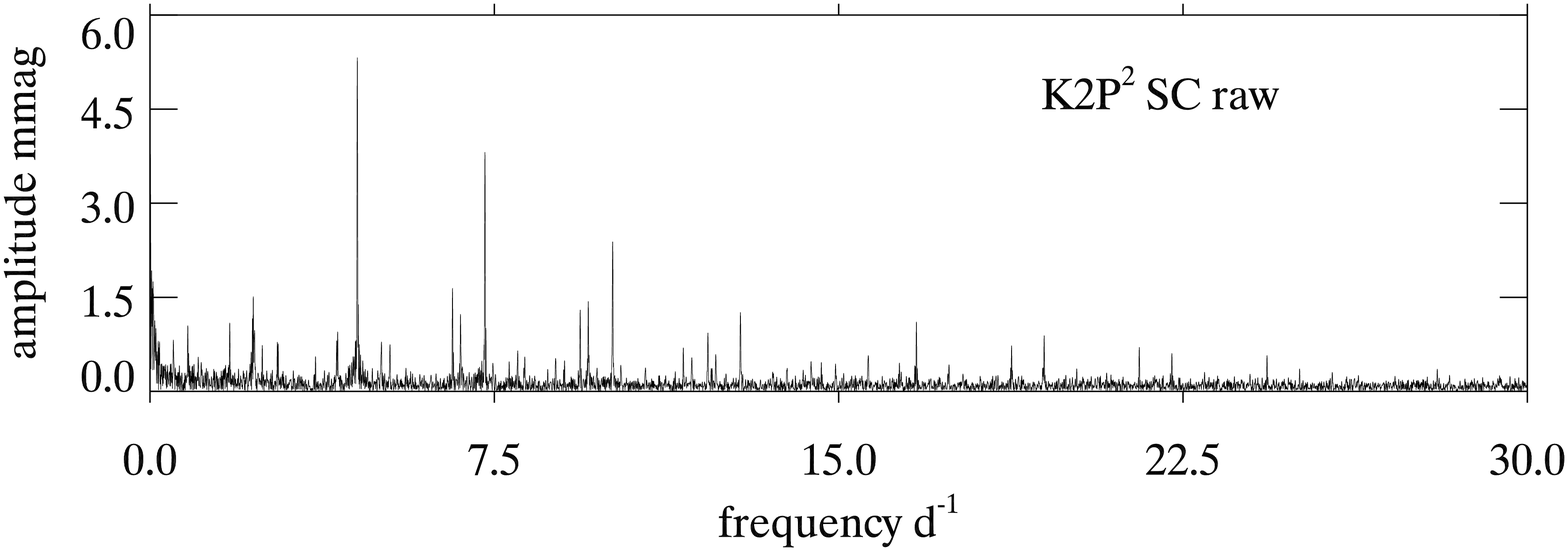}	
\includegraphics[width=0.49\linewidth,angle=0]{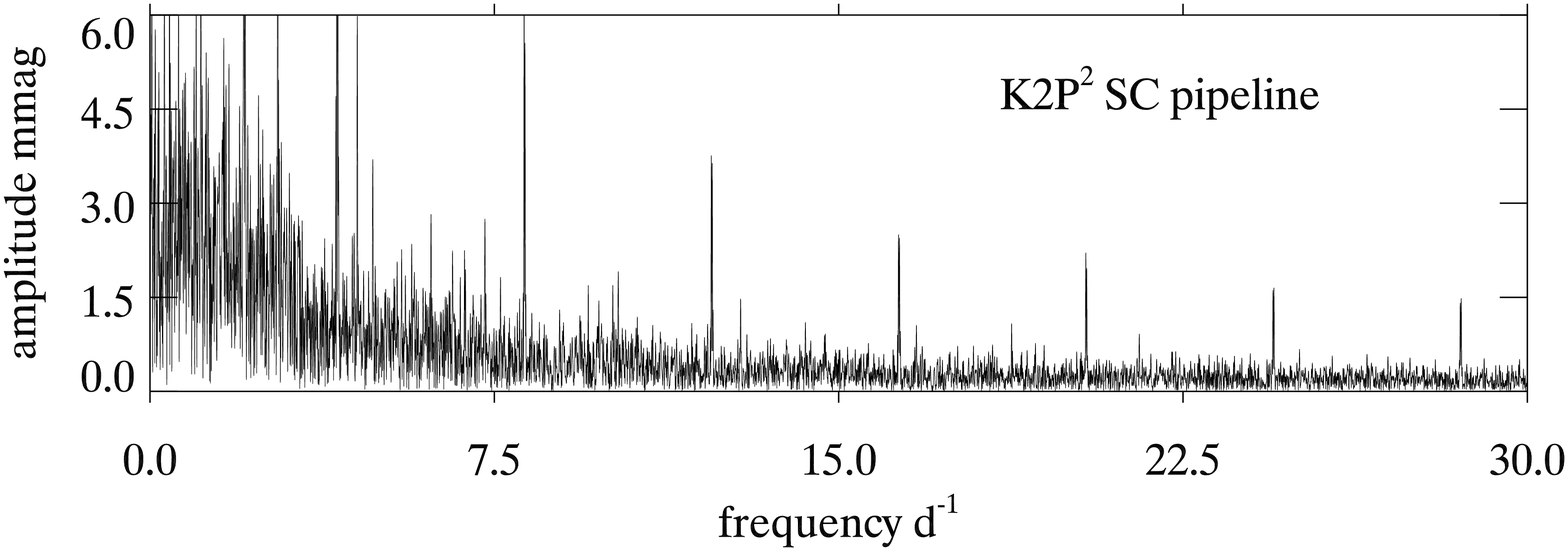}	
\includegraphics[width=0.49\linewidth,angle=0]{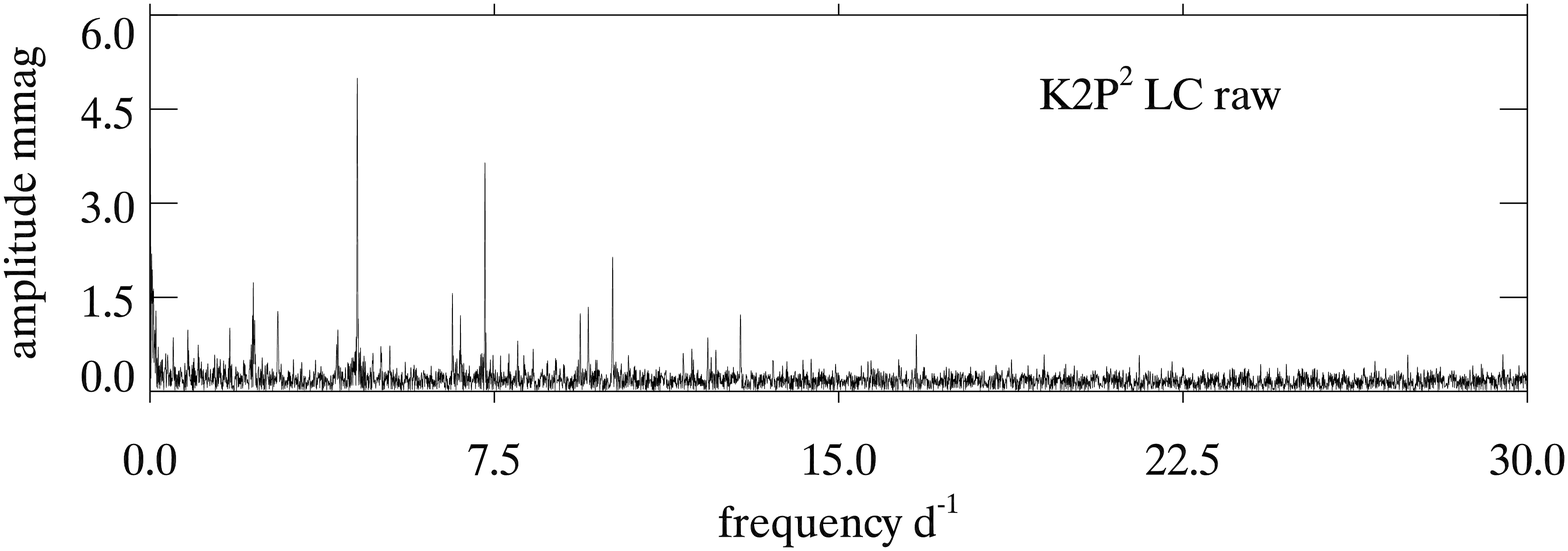}	
\includegraphics[width=0.49\linewidth,angle=0]{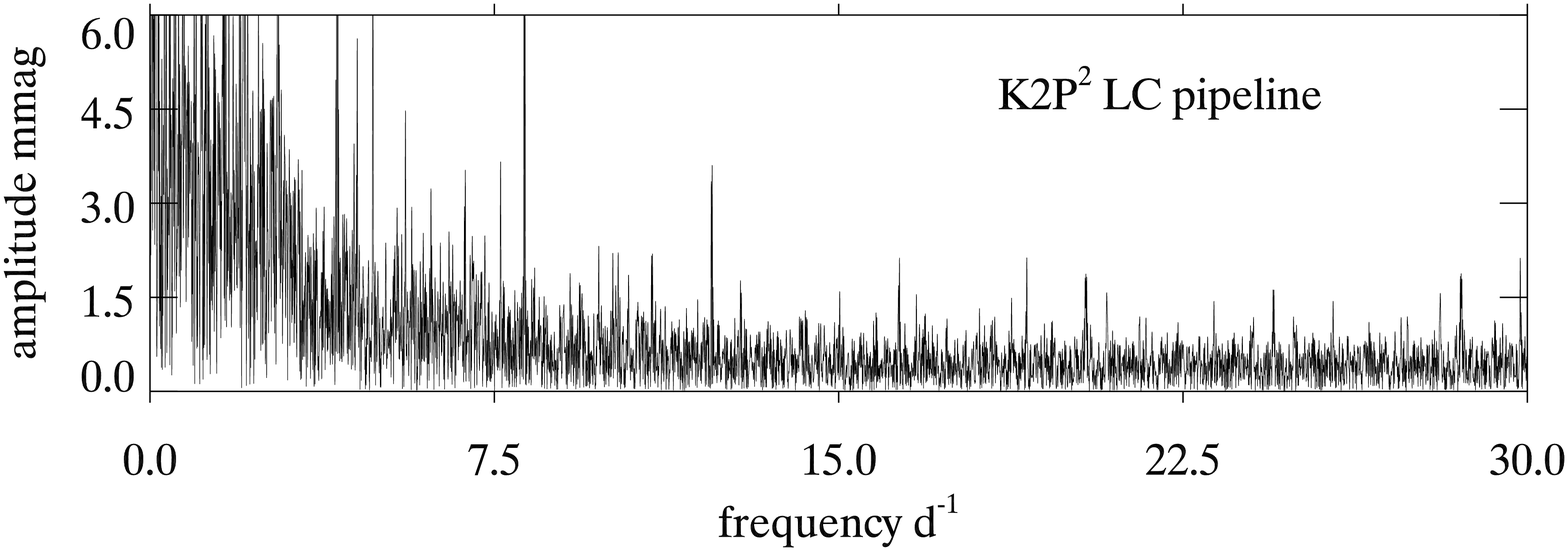}	
\includegraphics[width=0.49\linewidth,angle=0]{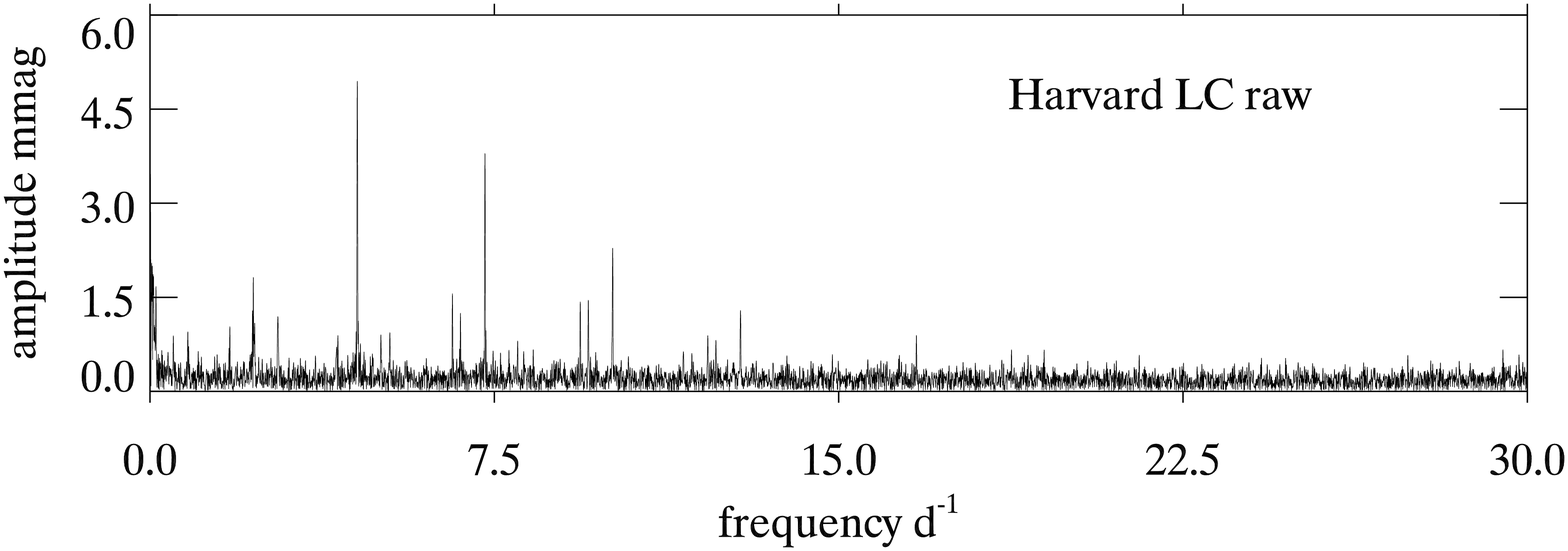}	
\includegraphics[width=0.49\linewidth,angle=0]{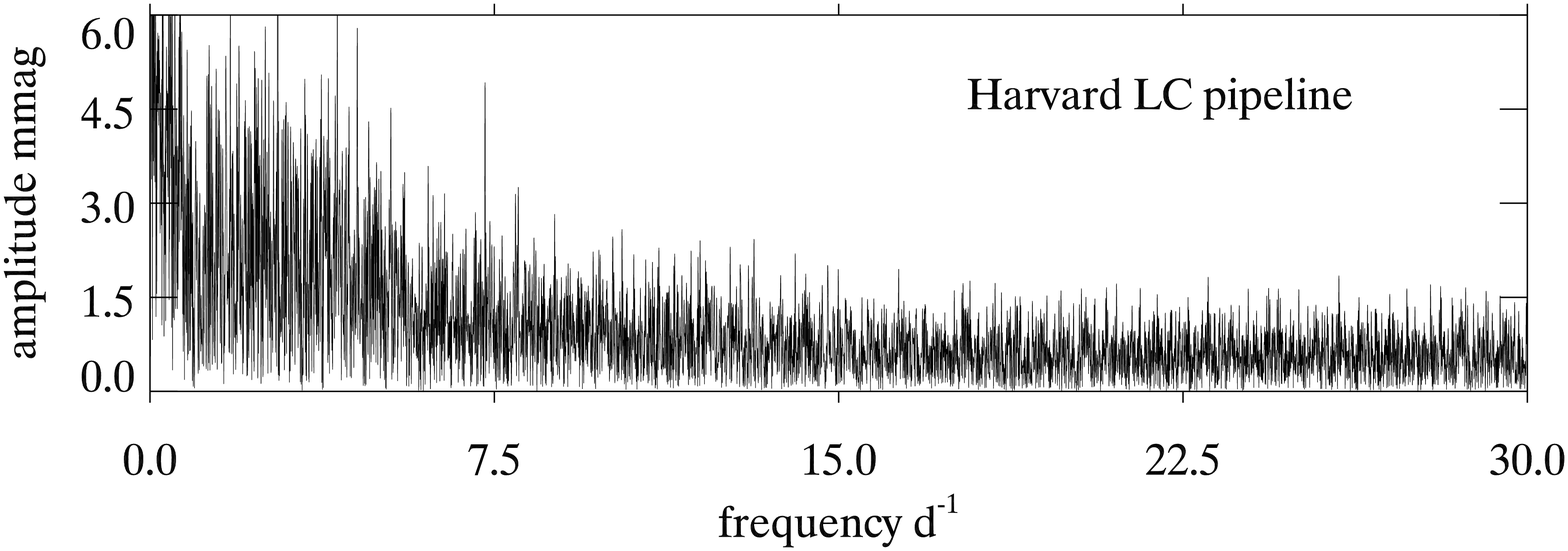}	
\includegraphics[width=0.49\linewidth,angle=0]{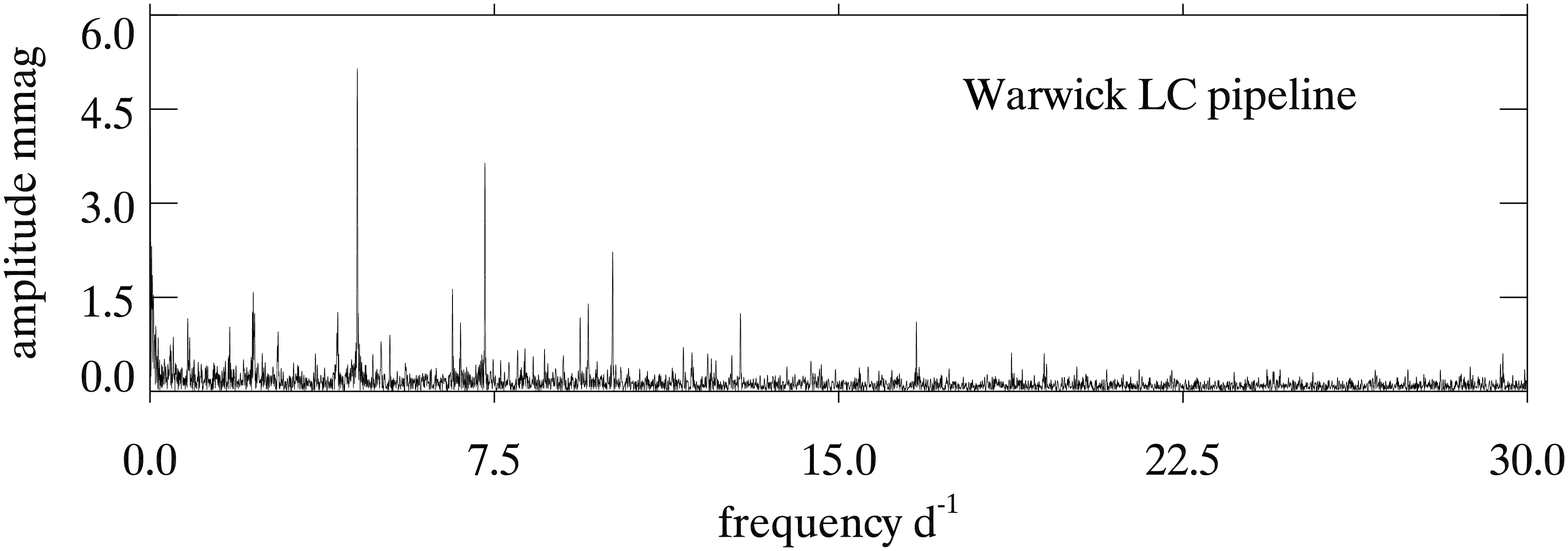}	
\caption{This diagram compares the amplitude spectra of the residuals to the 44-frequency fit given in Table\,\ref{tab:20158_2} for different data reductions. Left-hand column and bottom centre: these compare the raw data and the Warwick pipeline, which give similar results; Right-hand column and bottom centre: These compare the various pipeline reductions: K2P$^2$ SC and LC pipeline data (\citealt{handberg-lund2014}; \citealt{lund2015}); Harvard LC pipeline data \citep{vandenburg2014}; Warwick LC pipeline data \citep{armstrong2015}.  The highest peak is $\nu_3$; immediately to its left is the $<$1-mmag thruster firing artefact at 4.09\,d$^{-1}$, for comparison. The relatively high noise at low frequency in some of the pipeline data is a consequence of the high amplitude of the RR\,Lyr programme star; it is not indicative of the quality of the pipelines for more tractable light curves. }
\label{fig:pipelines}
\end{figure*}

We compared the amplitude spectra of the residuals to the 44-frequency fit for K2P$^2$ SC and LC raw and pipeline data (\citealt{handberg-lund2014}; \citealt{lund2015}), Harvard LC raw and pipeline data \citep{vandenburg2014} and Warwick LC pipeline data \citep{armstrong2015} with results seen in Fig.\,\ref{fig:pipelines}. It is clear that some pipeline data have increased low-frequency noise as a result of the time scale of the rise and fall of the RR\,Lyr light curve being close to the 5.9-hr thruster firing time scale, making it difficult to separate the stellar and instrumental variations. In this case, it is preferable not to model the pointing corrections. We also do not recommend a hands-on correction segment-by-segment, either by eye or with, say, polynomial fits, because this has as its basis a mental model that is necessarily subjective and not reproducible by other investigators. For the K2 RR\,Lyr stars we therefore recommend carefully chosen masks (either automatic or custom), flat fielding and outlier removal, but no corrections for the pointing changes; that is, the use of what we call `raw' data. 

\section{Conclusions}

We have discovered a new, rare triple-mode RR\,Lyr star, EPIC\,201585823, in the {\it Kepler} K2 mission Campaign 1 data. This star pulsates primarily in the fundamental and first-overtone radial modes, and, in addition, a third nonradial mode. The ratio of the period of the nonradial mode to that of the first-overtone radial mode, 0.616285, is remarkably similar to that seen in 11 other triple-mode RR\,Lyr stars, and in 260 RRc stars observed in the Galactic Bulge. There are 10 RRc stars observed from space \citep{moskalik2015} and 9 of them show the nonradial mode with the 0.61 period ratio. The number of similar RRc and RRd stars observed from the ground is growing rapidly. With further high precision space data and large-scale ground-based studies these behaviours will be seen to be normal in many RR\,Lyr stars. This systematic character promises new constraints on RR\,Lyr star models. 

We detected subharmonics of the nonradial mode frequency, which are a signature of period doubling of this oscillation; we note that this phenomenon is ubiquitous in RRc and RRd stars observed from space, and from ground with sufficient precision. The nonradial mode and subharmonic frequencies are not constant in frequency or in amplitude. 

The amplitude spectrum of EPIC\,201585823 is dominated by many combination frequencies among the three interacting pulsation mode frequencies. Inspection of the phase relationships of the combination frequencies in a phasor plot explains the `upward' shape of the light curve. 

We found that raw data with custom masks encompassing all pixels with significant signal for the star, but without correction for pointing changes, is best for frequency analysis of this star, and, by implication, other RR\,Lyr stars observed by the K2 mission.  

\section{acknowledgements}

Some/all of the data presented in this paper were obtained from the Mikulski Archive for Space Telescopes (MAST). STScI is operated by the Association of Universities for Research in Astronomy, Inc., under NASA contract NAS5-26555. Support for MAST for non-HST data is provided by the NASA Office of Space Science via grant NNX09AF08G and by other grants and contracts. This work has used data for EPIC\,201585823, which is one of the K2 targets selected and proposed by the RR\,Lyrae and Cepheid Working Group of the Kepler Asteroseismic Science Consortium (proposal number GO1067).  We thank Dr David Armstrong for discussion concerning the Warwick pipeline data, and Dr Simon Murphy for discussion concerning the phasor plots. DWK and DMB are funded by the UK STFC. SJE was supported during this research by a University of Central Lancashire Undergraduate Research Internship.  PM is supported by the Polish National Science Center through grant DEC-2012/05/B/ST9/03932. RH and MNL are supported by funding to the Stellar Astrophysics Centre at Aarhus University provided by the Danish National Research Foundation (Grant DNRF106), and by the ASTERISK project (ASTERoseismic Investigations with SONG and Kepler) funded by the European Research Council (Grant agreement no.: 267864). 

\bibliography{arXiv_201585823}

\begin{thebibliography}{40}
\expandafter\ifx\csname natexlab\endcsname\relax\def\natexlab#1{#1}\fi

\bibitem[{{Aerts}, {Christensen-Dalsgaard} \& {Kurtz}(2010){Aerts},
  {Christensen-Dalsgaard}, \& {Kurtz}}]{aertsetal2010}
{Aerts} C., {Christensen-Dalsgaard} J., {Kurtz} D.~W., 2010, {Asteroseismology,
  Astronomy and Astrophysics Library.~ISBN 978-1-4020-5178-4.~Springer
  Science+Business Media B.V.}

\bibitem[{{Armstrong} {et~al}\mbox{.}(2015){Armstrong}, {Kirk}, {Lam},
  {McCormac}, {Walker}, {Brown}, {Osborn}, {Pollacco}, \&
  {Spake}}]{armstrong2015}
{Armstrong} D.~J. {et~al.}, 2015, \aap, 579, A19

\bibitem[{{Bailey}(1902)}]{bailey1902}
{Bailey} S.~I., 1902, Annals of Harvard College Observatory, 38, 1

\bibitem[{{Bla{\v z}ko}(1907)}]{blazkho1907}
{Bla{\v z}ko} S., 1907, Astronomische Nachrichten, 175, 325

\bibitem[{{Bowman} \& {Kurtz}(2014)}]{bowman2014}
{Bowman} D.~M., {Kurtz} D.~W., 2014, \mnras, 444, 1909

\bibitem[{{Buchler} \& {Moskalik}(1992)}]{buchler1992}
{Buchler} J.~R., {Moskalik} P., 1992, \apj, 391, 736

\bibitem[{{Chadid}(2012)}]{chadid2012}
{Chadid} M., 2012, \aap, 540, A68

\bibitem[{{Gruberbauer} {et~al}\mbox{.}(2007){Gruberbauer}, {Kolenberg},
  {Rowe}, {Huber}, {Matthews}, {Reegen}, {Kuschnig}, {Cameron}, {Kallinger},
  {Weiss}, {Guenther}, {Moffat}, {Rucinski}, {Sasselov}, \&
  {Walker}}]{gruberbauer2007}
{Gruberbauer} M. {et~al.}, 2007, \mnras, 379, 1498

\bibitem[{{Handberg} \& {Lund}(2014)}]{handberg-lund2014}
{Handberg} R., {Lund} M.~N., 2014, \mnras, 445, 2698

\bibitem[{{Hertzsprung}(1913)}]{hertzsprung1913}
{Hertzsprung} E., 1913, Astronomische Nachrichten, 196, 201

\bibitem[{{Howell} {et~al}\mbox{.}(2014){Howell}, {Sobeck}, {Haas}, {Still},
  {Barclay}, {Mullally}, {Troeltzsch}, {Aigrain}, {Bryson}, {Caldwell},
  {Chaplin}, {Cochran}, {Huber}, {Marcy}, {Miglio}, {Najita}, {Smith},
  {Twicken}, \& {Fortney}}]{howell2014}
{Howell} S.~B. {et~al.}, 2014, \pasp, 126, 398

\bibitem[{{Jerzykiewicz} \& {Wenzel}(1977)}]{jerz1977}
{Jerzykiewicz} M., {Wenzel} W., 1977, Acta Ast., 27, 35

\bibitem[{{Jurcsik} {et~al}\mbox{.}(2015){Jurcsik}, {Smitola}, {Hajdu},
  {S{\'o}dor}, {Nuspl}, {Kolenberg}, {F{\H u}r{\'e}sz}, {Mo{\'o}r}, {Kun},
  {P{\'a}l}, {Bakos}, {Kelemen}, {Kov{\'a}cs}, {Kriskovics}, {S{\'a}rneczky},
  {Szalai}, {Szing}, \& {Vida}}]{jurcsik2015}
{Jurcsik} J. {et~al.}, 2015, \apjs, 219, 25

\bibitem[{{Koll{\'a}th}, {Moln{\'a}r} \& {Szab{\'o}}(2011){Koll{\'a}th},
  {Moln{\'a}r}, \& {Szab{\'o}}}]{kollath2011}
{Koll{\'a}th} Z., {Moln{\'a}r} L., {Szab{\'o}} R., 2011, \mnras, 414, 1111

\bibitem[{{Kurtz}(1985)}]{kurtz85}
{Kurtz} D.~W., 1985, \mnras, 213, 773

\bibitem[{{Kurtz} {et~al}\mbox{.}(2015){Kurtz}, {Shibahashi}, {Murphy},
  {Bedding}, \& {Bowman}}]{kurtzetal2015}
{Kurtz} D.~W., {Shibahashi} H., {Murphy} S.~J., {Bedding} T.~R., {Bowman}
  D.~M., 2015, \mnras, 450, 3015

\bibitem[{{Leavitt} \& {Pickering}(1912)}]{leavitt1912}
{Leavitt} H.~S., {Pickering} E.~C., 1912, Harvard College Observatory Circular,
  173, 1

\bibitem[{{Lenz} \& {Breger}(2004)}]{lenz&breger2004}
{Lenz} P., {Breger} M., 2004, in IAU Symposium, Vol. 224, The A-Star Puzzle,
  {J.~Zverko, J.~Ziznovsky, S.~J.~Adelman, \& W.~W.~Weiss}, ed., {CUP}, pp.
  786--790

\bibitem[{{Lund} {et~al}\mbox{.}(2015){Lund}, {Handberg}, {Davies}, {Chaplin},
  \& {Jones}}]{lund2015}
{Lund} M.~N., {Handberg} R., {Davies} G.~R., {Chaplin} W.~J., {Jones} C.~D.,
  2015, \apj, 806, 30

\bibitem[{{Moln{\'a}r} {et~al}\mbox{.}(2015){Moln{\'a}r}, {Szab{\'o}},
  {Moskalik}, {Nemec}, {Guggenberger}, {Smolec}, {Poleski}, {Plachy},
  {Kolenberg}, \& {Koll{\'a}th}}]{molnar2015}
{Moln{\'a}r} L. {et~al.}, 2015, \mnras, 452, 4283

\bibitem[{{Montgomery} \& {O'Donoghue}(1999)}]{montgomery-odonoghue99}
{Montgomery} M.~H., {O'Donoghue} D., 1999, Delta Scuti Star Newsletter, 13, 28

\bibitem[{{Moskalik}(2013)}]{moskalik2013}
{Moskalik} P., 2013, in Astrophysics and Space Science Proceedings, Vol.~31,
  Stellar Pulsations: Impact of New Instrumentation and New Insights,
  {Su{\'a}rez} J.~C., {Garrido} R., {Balona} L.~A., {Christensen-Dalsgaard} J.,
  eds., p. 103

\bibitem[{{Moskalik}(2014)}]{moskalik2014}
---, 2014, in IAU Symposium, Vol. 301, IAU Symposium, {Guzik} J.~A., {Chaplin}
  W.~J., {Handler} G., {Pigulski} A., eds., pp. 249--256

\bibitem[{{Moskalik} \& {Buchler}(1990)}]{moskalik1990}
{Moskalik} P., {Buchler} J.~R., 1990, \apj, 355, 590

\bibitem[{{Moskalik} {et~al}\mbox{.}(2015){Moskalik}, {Smolec}, {Kolenberg},
  {Moln{\'a}r}, {Kurtz}, {Szab{\'o}}, {Benk{\H o}}, {Nemec}, {Chadid},
  {Guggenberger}, {Ngeow}, {Jeon}, {Kopacki}, \& {Kanbur}}]{moskalik2015}
{Moskalik} P. {et~al.}, 2015, \mnras, 447, 2348

\bibitem[{{Netzel}, {Smolec} \& {Moskalik}(2015{\natexlab{a}}){Netzel},
  {Smolec}, \& {Moskalik}}]{netzel2015a}
{Netzel} H., {Smolec} R., {Moskalik} P., 2015{\natexlab{a}}, \mnras, 447, 1173

\bibitem[{{Netzel}, {Smolec} \& {Moskalik}(2015{\natexlab{b}}){Netzel},
  {Smolec}, \& {Moskalik}}]{netzel2015b}
---, 2015{\natexlab{b}}, \mnras, 453, 2022

\bibitem[{{Olech} \& {Moskalik}(2009)}]{olech2009}
{Olech} A., {Moskalik} P., 2009, \aap, 494, L17

\bibitem[{{Smolec} {et~al}\mbox{.}(2012){Smolec}, {Soszy{\'n}ski}, {Moskalik},
  {Udalski}, {Szyma{\'n}ski}, {Kubiak}, {Pietrzy{\'n}ski}, {Wyrzykowski},
  {Ulaczyk}, {Poleski}, {Koz{\l}owski}, \& {Pietrukowicz}}]{smolec2012}
{Smolec} R. {et~al.}, 2012, \mnras, 419, 2407

\bibitem[{{Smolec} {et~al}\mbox{.}(2015{\natexlab{a}}){Smolec},
  {Soszy{\'n}ski}, {Udalski}, {Szyma{\'n}ski}, {Pietrukowicz}, {Skowron},
  {Koz{\l}owski}, {Poleski}, {Moskalik}, {Skowron}, {Pietrzy{\'n}ski},
  {Wyrzykowski}, {Ulaczyk}, \& {Mr{\'o}z}}]{smolecetal2015b}
---, 2015{\natexlab{a}}, \mnras, 447, 3873

\bibitem[{{Smolec} {et~al}\mbox{.}(2015{\natexlab{b}}){Smolec},
  {Soszy{\'n}ski}, {Udalski}, {Szyma{\'n}ski}, {Pietrukowicz}, {Skowron},
  {Koz{\l}owski}, {Poleski}, {Skowron}, {Pietrzy{\'n}ski}, {Wyrzykowski},
  {Ulaczyk}, \& {Mr{\'o}z}}]{smolecetal2015a}
---, 2015{\natexlab{b}}, \mnras, 447, 3756

\bibitem[{{Soszy{\'n}ski} {et~al}\mbox{.}(2011){Soszy{\'n}ski}, {Dziembowski},
  {Udalski}, {Poleski}, {Szyma{\'n}ski}, {Kubiak}, {Pietrzy{\'n}ski},
  {Wyrzykowski}, {Ulaczyk}, {Koz{\l}owski}, \& {Pietrukowicz}}]{soszynski2011}
{Soszy{\'n}ski} I. {et~al.}, 2011, Acta Astron., 61, 1

\bibitem[{{Soszy{\'n}ski} {et~al}\mbox{.}(2014){Soszy{\'n}ski}, {Udalski},
  {Szyma{\'n}ski}, {Pietrukowicz}, {Mr{\'o}z}, {Skowron}, {Koz{\l}owski},
  {Poleski}, {Skowron}, {Pietrzy{\'n}ski}, {Wyrzykowski}, {Ulaczyk}, \&
  {Kubiak}}]{soszynski2014}
---, 2014, Acta Astron., 64, 177

\bibitem[{{Szab{\'o}} {et~al}\mbox{.}(2014){Szab{\'o}}, {Benk{\H o}},
  {Papar{\'o}}, {Chapellier}, {Poretti}, {Baglin}, {Weiss}, {Kolenberg},
  {Guggenberger}, \& {Le Borgne}}]{szabo2014}
{Szab{\'o}} R. {et~al.}, 2014, \aap, 570, A100

\bibitem[{{Szab{\'o}} {et~al}\mbox{.}(2010){Szab{\'o}}, {Koll{\'a}th},
  {Moln{\'a}r}, {Kolenberg}, {Kurtz}, {Bryson}, {Benk{\H o}},
  {Christensen-Dalsgaard}, {Kjeldsen}, {Borucki}, {Koch}, {Twicken}, {Chadid},
  {di Criscienzo}, {Jeon}, {Moskalik}, {Nemec}, \& {Nuspl}}]{szabo2010}
---, 2010, \mnras, 409, 1244

\bibitem[{{Tsesevich}(1953)}]{tsessivich1953}
{Tsesevich} V.~P., 1953, Trudy Gosudarstvennogo Astronomicheskogo Instituta,
  23, 62

\bibitem[{{Udalski} {et~al}\mbox{.}(2008){Udalski}, {Szyma{\'n}ski},
  {Soszy{\'n}ski}, \& {Poleski}}]{udalski2008}
{Udalski} A., {Szyma{\'n}ski} M.~K., {Soszy{\'n}ski} I., {Poleski} R., 2008,
  Acta Astron., 58, 69

\bibitem[{{Udalski}, {Szyma{\'n}ski} \& {Szyma{\'n}ski}(2015){Udalski},
  {Szyma{\'n}ski}, \& {Szyma{\'n}ski}}]{udalski2015}
{Udalski} A., {Szyma{\'n}ski} M.~K., {Szyma{\'n}ski} G., 2015, Acta Astron.,
  65, 1

\bibitem[{{Vanderburg} \& {Johnson}(2014)}]{vandenburg2014}
{Vanderburg} A., {Johnson} J.~A., 2014, \pasp, 126, 948

\bibitem[{{Wallerstein}(2002)}]{wallerstein2002}
{Wallerstein} G., 2002, \pasp, 114, 689

\end{thebibliography}

\end{document}